\documentclass[aip,graphicx]{revtex4-1}
\usepackage[OT2,T1]{fontenc}
\usepackage[russian,USenglish]{babel}
\draft % marks overfull lines with a black rule on the right

\begin{document}

\title{On the nature of the new group LB1} %Title of paper

\author{Alcides Garat}
%\thanks{}
\affiliation{Former Professor at Universidad de la Rep\'{u}blica, Av. 18 de Julio 1824-1850, 11200 Montevideo, Uruguay.}
\date{\today}
%\date{November 9th, 2011}

\begin{abstract}
The new local group LB1 introduced previously will be studied and reviewed in detail, depicting its unique nature that makes it a new group in fundamental physics. It will be made clear that even though most of its elements are Lorentz transformations, one unique discrete transformation not present in the Lorentz groups, is making this group into a new group because it is a reflection. In addition there will be four particular transformations onto the local light cone. It is these discrete transformations that allow for an isomorphism between the group $U(1)$ and LB1. This result will have profound resonations in all of particle physics, general relativity, relativistic astrophysics, Riemannian geometry and group theory. These new group will be associated to a whole set of new experiments put forward.
\end{abstract}

\keywords{Einstein-Maxwell four-dimensional Lorentzian spacetimes; new tetrads; new groups; new groups isomorphisms; singular gauge; non-null electromagnetic fields}
\pacs{12.10.-g; 04.40.Nr; 04.20.Cv; 11.15.-q; 02.40.Ky; 02.20.Qs\\ MSC2010: 51H25; 53c50; 20F65; 70s15; 70G65; 70G45}% insert suggested PACS

%\pacs{}% insert suggested PACS numbers in braces on next line

\maketitle %\maketitle must follow title, authors, abstract and \pacs

\section{Introduction}
\label{intro}

The purpose of this manuscript is to make clear the nature of the group LB1 through a detailed study of its elements. To these end we will consider just the fundamental steps associated to tetrad construction in a four-dimensional Lorentzian curved spacetime that lead to our group results. The whole analysis as given in manuscripts \cite{A,ROMP,SING,ATGU,GLAW,TSTRUCT,AEO,AMONO} will be reviewed. The goal of this review will be to create an encompassing compendium as thorough as possible about this new subject of research that has been developed in the past twenty five years. We will in the space allocated intend to explain the basic fundamental elements that make up the basis of this area of geometry and display several applications with the aim of highlighting the relevance of this topic. Therefore, we will start this work introducing the tetrads found in the aforementioned paper \cite{A} that locally and covariantly diagonalize the electromagnetic stress-energy tensor. This local tetrad electromagnetic gauge transformations reflect the existence of an isomorphism between the local internal group $U(1)$ and the local group of spacetime transformations LB1. The timelike vector and one spacelike vector span the local plane one while the other two spacelike vectors span the local orthogonal plane two. The local group of electromagnetic gauge transformations is isomorphic on the local plane one to the local group LB1. The group LB1 is given by $SO(1,1) \times Z_{2} \times Z_{2}$ where $SO(1,1)$ is proper orthochronous. The first $Z_{2}$ is given by $\{I_{2 \times 2}, -I_{2 \times 2}\}$ and the second $Z_{2}$ is given by $\{I_{2 \times 2}, \mbox{the swap}\: (01|10)\}$. There are two discrete transformations. One of them that we designated as $-I_{2 \times 2}$ is the full inversion two by two and the other a reflection designated as $\mbox{the swap}\: (01|10)$ and given by $\Lambda^{o}_{\:\:o} = 0$, $\Lambda^{o}_{\:\:1} = 1$, $\Lambda^{1}_{\:\:o} = 1$,  $\Lambda^{1}_{\:\:1} = 0$ which is not a Lorentz transformation, because it is a reflection. We would have to add in order to complete the image of the map $SO(1,1) \times Z_{2} \times Z_{2}\: \bigoplus \: \{light\:cone\:gauge\}$ where the light cone gauge includes the inhomogeneous two solutions to the differential equations in the local future and past light cones established in reference \cite{SING} where the reflection through the asymptote $Y=X$ will produce two more identical inhomogeneous solutions. A total of four and this analysis can be also found in sections \ref{timelikenull} and \ref{sec:appIV}. When we talk about the light cone gauge we are not referring to this concept in the sense of book \cite{MK} or manuscript \cite{JSJHS}. Light cone gauge means that when we map the tetrad vectors in the local plane one through specific local electromagnetic gauge transformations, the vectors that result of these transformations are null vectors. In manuscript \cite{SING} the general case has been presented as well as examples in the Coulomb and Reissner Nordstr\"{o}m geometries, see section \ref{timelikenull}. On the local orthogonal plane two and independently from the mapping in the plane one, the local group of electromagnetic gauge transformations is mapped onto the local group of spatial tetrad rotations $SO(2)$, see section \ref{sec:appIII}. By means of these tetrad vectors we will proceed to the study of their local gauge transformation properties and analyze one by one all the group elements in section \ref{localbladeone}. New tetrads were introduced in references \cite{A,ROMP,SING,ATGU,GLAW,TSTRUCT,AEO,AMONO} and \cite{MW} such that these set of four vectors diagonalizes in a local and covariant way, any electromagnetic stress-energy tensor for any non-null electromagnetic field in a curved four-dimensional Lorentzian spacetime. As mentioned above at every point in a curved Einstein-Maxwell spacetime the timelike and one spacelike tetrad vectors define plane or blade one \cite{SCH}. The other two spacelike vectors define plane or blade two, orthogonal to plane one. These tetrad vectors are built out of two construction structures. On one hand the skeleton, which is invariant under electromagnetic gauge transformations, and on the other hand the gauge vectors, that have a gauge dependence. Gauge vectors are spacetime gauge by themselves and they contain gauge in their construction. When we study the local gauge transformations of these tetrad vectors we observe the following. The two unit orthogonal vectors that locally define plane one, remain on plane one after the transformation, and similar for plane two. This property ensures the metric tensor invariance in a manifest fashion. These findings will have profound consequences in all of particle physics, general relativity, relativistic astrophysics, Riemannian geometry and group theory. These results stand at the crossroads of field unification and group unification simultaneously. Since this a review manuscript and we make use of the concept of local gauge and local gauge transformations, we will refer the reader to a number of reviews, books  and papers about the subject \cite{LGT1}$^{-}$\cite{LGT13}, where a substantial amount of key references on this topic can be found. Throughout the paper we use the conventions of manuscript \cite{MW}. We use a metric with sign conventions $-+++$. If $F_{\mu\nu}$ is the electromagnetic field then $f_{\mu\nu}= (G^{1/2} / c^2) \: F_{\mu\nu}$ is the geometrized electromagnetic field.

\section{Local gauge transformations on blade one}
\label{localbladeone}

Let us start by introducing the Einstein-Maxwell equations that can be written,

\begin{eqnarray}
f^{\mu\nu}_{\:\:\:\:\:;\nu} &=& 0 \label{EM1}\\
\ast f^{\mu\nu}_{\:\:\:\:\:;\nu} &=& 0 \label{EM2}\\
R_{\mu\nu} &=& f_{\mu\lambda}\:\:f_{\nu}^{\:\:\:\lambda}
+ \ast f_{\mu\lambda}\:\ast f_{\nu}^{\:\:\:\lambda}\ . \label{EM3}
\end{eqnarray}

In geometrodynamics, the Maxwell equations, $f^{\mu\nu}_{\:\:\:\:\:;\nu} = 0$ and $\ast f^{\mu\nu}_{\:\:\:\:\:;\nu} = 0$ are telling us that two potential vector fields $A_{\nu}$ and $\ast A_{\nu}$ exist \cite{CF}. Note that the star in $\ast A_{\nu}$ is just a name, not the dual Hodge operator, meaning that $\ast A_{\nu ;\mu} = (\ast A_{\nu})_{;\mu}$. The symbol $``;''$ stands for covariant derivative with respect to the metric tensor $g_{\mu\nu}$. When we make the transformation,

\begin{eqnarray}
A_{\alpha} \rightarrow A_{\alpha} + \Lambda_{,\alpha}\ , \label{2G1}
\end{eqnarray}

$f_{\mu\nu} = A_{\nu ;\mu} - A_{\mu ;\nu}$ remains invariant, and the transformation,

\begin{eqnarray}
\ast A_{\alpha} \rightarrow \ast A_{\alpha} + \ast \Lambda_{,\alpha}\ , \label{2G2}
\end{eqnarray}

leaves $\ast f_{\mu\nu} = \ast A_{\nu ;\mu} - \ast A_{\mu ;\nu}$ invariant, as long as the functions $\Lambda$ and $\ast \Lambda$ are local
scalars. These two potentials are not independent from each other, but necessary in the construction of the new local tetrad that we will introduce shortly \cite{Addendum A}. We briefly remind ourselves that the original expression for the electromagnetic stress-energy tensor
$T_{\mu\nu}= f_{\mu\lambda}\:\:f_{\nu}^{\:\:\:\lambda} + \ast f_{\mu\lambda}\:\ast f_{\nu}^{\:\:\:\lambda}$ was given in terms of the electromagnetic tensor $f_{\mu\nu}$ and its dual $\ast f_{\mu\nu}={1 \over 2}\:\epsilon_{\mu\nu\sigma\tau}\:f^{\sigma\tau}$. The alternating tensor $\epsilon_{\mu\nu\sigma\tau}$ is explained in section \ref{sec:appI}.

After a local duality transformation,

\begin{equation}
f_{\mu\nu} = \xi_{\mu\nu} \: \cos\alpha + \ast\xi_{\mu\nu} \: \sin\alpha\ ,\label{2dr}
\end{equation}

where the local scalar $\alpha$ is the complexion, we are able to write the stress-energy in terms of the extremal field $\xi_{\mu\nu}$ and its dual as $T_{\mu\nu}=\xi_{\mu\lambda}\:\:\xi_{\nu}^{\:\:\:\lambda} + \ast \xi_{\mu\lambda}\:\ast \xi_{\nu}^{\:\:\:\lambda}$. We can express the extremal field as,

\begin{equation}
\xi_{\mu\nu} = e^{-\ast \alpha} f_{\mu\nu}\ = \cos\alpha\:f_{\mu\nu} - \sin\alpha\:\ast f_{\mu\nu}.\label{2dref}
\end{equation}

Extremal fields are local gauge invariants in the electromagnetic sense as it can be noticed from equation (\ref{2dref}). Extremal fields satisfy the equation

\begin{equation}
\xi_{\mu\nu} \ast \xi^{\mu\nu}= 0\ . \label{2i0}
\end{equation}

This a condition imposed on extremal fields in order to find a local scalar named the complexion $\alpha$. The explicit expression for the complexion can be given when imposing condition (\ref{2i0}) on equation (\ref{2dref}) by $\tan(2\alpha) = - f_{\mu\nu}\:\ast f^{\mu\nu} / f_{\lambda\rho}\:f^{\lambda\rho}$. It can be proved that condition (\ref{2i0}) and through the use of the general identity,

\begin{eqnarray}
A_{\mu\alpha}\:B^{\nu\alpha} -
\ast B_{\mu\alpha}\: \ast A^{\nu\alpha} &=& \frac{1}{2}
\: \delta_{\mu}^{\:\:\:\nu}\: A_{\alpha\beta}\:B^{\alpha\beta}  \ ,\label{2ig}
\end{eqnarray}

which is valid for every pair of antisymmetric tensors in a four-dimensional Lorentzian spacetime \cite{MW}, when applied to the case
$A_{\mu\alpha} = \xi_{\mu\alpha}$ and $B^{\nu\alpha} = \ast \xi^{\nu\alpha}$ yields the equivalent condition,

\begin{eqnarray}
\xi_{\alpha\mu}\:\ast \xi^{\mu\nu} &=& 0\ .\label{2i2}
\end{eqnarray}

When applying equation (\ref{2ig}) to the case $A_{\mu\alpha}=\xi_{\mu\alpha}$ and $B^{\nu\alpha}=\xi^{\nu\alpha}$ yields the identity,

\begin{eqnarray}
\xi_{\mu\alpha}\:\xi^{\nu\alpha} - \ast \xi_{\mu\alpha}\: \ast \xi^{\nu\alpha} &=& \frac{1}{2} \: \delta_{\mu}^{\:\:\:\nu}\ Q \ ,\label{2i3}
\end{eqnarray}

where $Q=\xi_{\mu\nu}\:\xi^{\mu\nu}$. There are four tetrad vectors that at every point in spacetime diagonalize the stress-energy tensor $T_{\mu\nu}=\xi_{\mu\lambda}\:\:\xi_{\nu}^{\:\:\:\lambda} + \ast \xi_{\mu\lambda}\:\ast \xi_{\nu}^{\:\:\:\lambda}$ in geometrodynamics,

\begin{eqnarray}
V_{(1)general}^{\alpha} &=& \xi^{\alpha\lambda}\:\xi_{\rho\lambda}\:X^{\rho}
\label{V1}\\
V_{(2)general}^{\alpha} &=& \sqrt{-Q/2} \: \xi^{\alpha\lambda} \: X_{\lambda}
\label{V2}\\
V_{(3)general}^{\alpha} &=& \sqrt{-Q/2} \: \ast \xi^{\alpha\lambda} \: Y_{\lambda}
\label{V3}\\
V_{(4)general}^{\alpha} &=& \ast \xi^{\alpha\lambda}\: \ast \xi_{\rho\lambda}
\:Y^{\rho}\ .\label{V4}
\end{eqnarray}

These tetrad vectors are all eigenvectors of the stress-energy tensor. We can test this assertion when using iteratively equations (\ref{2i2}) and (\ref{2i3}). In geometrodynamics, the Maxwell equations,

\begin{eqnarray}
f^{\mu\nu}_{\:\:\:\:\:;\nu} &=& 0 \label{L1}\nonumber\\
\ast f^{\mu\nu}_{\:\:\:\:\:;\nu} &=& 0 \ , \label{L2}
\end{eqnarray}

reveal that two potential vector fields $A_{\nu}$ and $\ast A_{\nu}$ exist \cite{CF},

\begin{eqnarray}
f_{\mu\nu} &=& A_{\nu ;\mu} - A_{\mu ;\nu}\label{ER}\nonumber\\
\ast f_{\mu\nu} &=& \ast A_{\nu ;\mu} - \ast A_{\mu ;\nu} \ .\label{DER}
\end{eqnarray}

We have therefore the possible choice $X^{\rho}=A^{\rho}$ and $Y^{\rho}=\ast A^{\rho}$, please see reference \cite{Addendum A} regarding the nature of these two potentials and other possible choices for the gauge vectors $X^{\rho}$ and $Y^{\rho}$. With all these elements we proceed to introduce without further delay the new orthonormal tetrad that diagonalizes locally and covariantly the stress-energy tensor,

\begin{eqnarray}
U^{\alpha} &=& \xi^{\alpha\lambda}\:\xi_{\rho\lambda}\:A^{\rho} \:
/ \: (\: \sqrt{-Q/2} \: \sqrt{A_{\mu} \ \xi^{\mu\sigma} \
\xi_{\nu\sigma} \ A^{\nu}}\:) \label{2U}\\
V^{\alpha} &=& \xi^{\alpha\lambda}\:A_{\lambda} \:
/ \: (\:\sqrt{A_{\mu} \ \xi^{\mu\sigma} \
\xi_{\nu\sigma} \ A^{\nu}}\:) \label{2V}\\
Z^{\alpha} &=& \ast \xi^{\alpha\lambda} \: \ast A_{\lambda} \:
/ \: (\:\sqrt{\ast A_{\mu}  \ast \xi^{\mu\sigma}
\ast \xi_{\nu\sigma}  \ast A^{\nu}}\:)
\label{2Z}\\
W^{\alpha} &=& \ast \xi^{\alpha\lambda}\: \ast \xi_{\rho\lambda}
\:\ast A^{\rho} \: / \: (\:\sqrt{-Q/2} \: \sqrt{\ast A_{\mu}
\ast \xi^{\mu\sigma} \ast \xi_{\nu\sigma} \ast A^{\nu}}\:) \ .
\label{2W}
\end{eqnarray}

where $Q=\xi_{\mu\nu}\:\xi^{\mu\nu}=-\sqrt{T_{\mu\nu}T^{\mu\nu}}$ according to equations (39) in \cite{MW}. $Q$ is assumed not to be zero,
because we are dealing with non-null electromagnetic fields. Non-null we clarify means basically that $f_{\mu\nu}\:f^{\mu\nu}\neq0$ and $\ast f_{\mu\nu}\:f^{\mu\nu}\neq0$. In turn and by definitions these last equations imply that $\xi_{\mu\nu}\:\xi^{\mu\nu}\neq0$. The first two (\ref{2U}-\ref{2V}) are eigenvectors of the stress-energy tensor with eigenvalue $Q/2$, the last two (\ref{2Z}-\ref{2W}) with eigenvalue $-Q/2$. Schouten defined what he called a two-bladed structure in a spacetime \cite{SCH}. Tetrad vectors (\ref{2U}-\ref{2V}) define the local blade one and vectors (\ref{2Z}-\ref{2W}) define the local blade two. Let us introduce some names. Setting aside normalization factors, the tetrad vectors have two essential components, see reference \cite{A}. For instance in vector $U^{\alpha}_{not-normalized} = \xi^{\alpha\lambda}\:\xi_{\rho\lambda}\:X^{\rho}$ there are two main structures. First, the skeleton, in this case $\xi^{\alpha\lambda}\:\xi_{\rho\lambda}$, and second, the gauge vector $X^{\rho}$. In vector $Z^{\alpha}_{not-normalized} = \ast \xi^{\alpha\lambda} \: Y_{\lambda}$, the skeleton is $\ast \xi^{\alpha\lambda}$, and the gauge vector is $Y^{\rho}$. The gauge vectors it was proved in manuscript \cite{A} could be anything that does not make the tetrad vectors trivial. That is, the tetrad (\ref{2U}-\ref{2W}) diagonalizes the stress-energy tensor for any non-trivial gauge vectors $X^{\mu}$ and $Y^{\mu}$. It was therefore proved that we can make different choices for $X^{\mu}$ and $Y^{\mu}$. We made the choices $X^{\mu} = A^{\mu}$ and $Y^{\mu} = \ast A^{\mu}$ and please see references \cite{Addendum,Addendum A} for other possible choices. Let us remember that the star in $\ast A^{\mu}$ is just a name. For the purpose of making the notation compatible with that of manuscript \cite{A}, let us introduce the non-normalized local tetrad that covariantly diagonalizes the electromagnetic stress-energy tensor,

\begin{eqnarray}
V_{(1)}^{\alpha} &=& \xi^{\alpha\lambda}\:\xi_{\rho\lambda}\:A^{\rho}
\label{2V1}\\
V_{(2)}^{\alpha} &=& \sqrt{-Q/2} \: \xi^{\alpha\lambda} \: A_{\lambda}
\label{2V2}\\
V_{(3)}^{\alpha} &=& \sqrt{-Q/2} \: \ast \xi^{\alpha\lambda} \:\ast A_{\lambda}
\label{2V3}\\
V_{(4)}^{\alpha} &=& \ast \xi^{\alpha\lambda}\: \ast \xi_{\rho\lambda}
\:\ast A^{\rho}\ ,\label{2V4}
\end{eqnarray}

We will use this particular version in order to study the tetrad local electromagnetic gauge transformations. Vector $V_{(1)}^{\alpha}$ is assumed for simplicity to be timelike, see reference \cite{A}. We reiterate that without repeating all of the analysis in manuscript \cite{A} we proceed to study the different possible local tetrad transformation cases on blade one resorting to the same notation employed in this previous work. In order to simplify the notation we will write $\Lambda_{,\alpha}=\Lambda_{\alpha}$.

The purpose of this work is to study the different cases that arise when we consider the change in (\ref{2V1}-\ref{2V2}) under (\ref{2G1}),

\begin{eqnarray}
\tilde{V}_{(1)}^{\alpha} &=& V_{(1)}^{\alpha} +
\xi^{\alpha\lambda}\:\xi_{\rho\lambda}\:\Lambda^{\rho}\label{2TU}\\
\tilde{V}_{(2)}^{\alpha} &=& V_{(2)}^{\alpha} +
\sqrt{-Q/2}\:\xi^{\alpha\lambda}\:\Lambda_{\lambda}\ ,\label{2TV}
\end{eqnarray}

Since the second terms in (\ref{2TU}-\ref{2TV}) and according to equation (\ref{2i2}) belong in the local plane or blade one,
we proceed then to write equations (\ref{2TU}-\ref{2TV}) as,

\begin{eqnarray}
\tilde{V}_{(1)}^{\alpha} &=& V_{(1)}^{\alpha} +
C\:V_{(1)}^{\alpha} + D\:V_{(2)}^{\alpha}\label{2TUN}\\
\tilde{V}_{(2)}^{\alpha} &=& V_{(2)}^{\alpha} +
E\:V_{(1)}^{\alpha} + F\:V_{(2)}^{\alpha}\ .\label{2TVN}
\end{eqnarray}

After some algebraic work it was found in reference \cite{A} the following relations between coefficients,

\begin{eqnarray}
E &=& D \label{2E=D}\\
F &=& C \label{2F=C} \ .
\end{eqnarray}

It was also found that,

\begin{eqnarray}
C&=&(-Q/2)\:V_{(1)\sigma}\:\Lambda^{\sigma} / (\:V_{(2)\beta}\:
V_{(2)}^{\beta}\:)\label{2COEFFC}\\
D&=&(-Q/2)\:V_{(2)\sigma}\:\Lambda^{\sigma} / (\:V_{(1)\beta}\:
V_{(1)}^{\beta}\:)\ .\label{2COEFFD}
\end{eqnarray}

After all this algebraic work we would like to calculate the norm of the transformed vectors $\tilde{V}_{(1)}^{\alpha}$ and $\tilde{V}_{(2)}^{\alpha}$,

\begin{eqnarray}
\tilde{V}_{(1)}^{\alpha}\:\tilde{V}_{(1)\alpha} &=&
[(1+C)^2-D^2]\:V_{(1)}^{\alpha}\:V_{(1)\alpha}\label{2FP}\\
\tilde{V}_{(2)}^{\alpha}\:\tilde{V}_{(2)\alpha} &=&
[(1+C)^2-D^2]\:V_{(2)}^{\alpha}\:V_{(2)\alpha}\ ,\label{2SP}
\end{eqnarray}

where the relation $V_{(1)}^{\alpha}\:V_{(1)\alpha} = -V_{(2)}^{\alpha}\:V_{(2)\alpha}$ has been used. It is evident from equations (\ref{2FP}-\ref{2SP})
that two situations might arise. Either $[(1+C)^2-D^2]$ is positive, or negative. Equality to zero will be analyzed at the end. The condition for these
transformations to keep the timelike or spacelike character of $V_{(1)}^{\alpha}$ and $V_{(2)}^{\alpha}$ is $[(1+C)^2-D^2]>0$. If this condition
is satisfied, then we can normalize the transformed vectors
$\tilde{V}_{(1)}^{\alpha}$ and $\tilde{V}_{(2)}^{\alpha}$ in expressions (\ref{2TUN}-\ref{2TVN}) as follows,

\begin{eqnarray}
{\tilde{V}_{(1)}^{\alpha}
\over \sqrt{-\tilde{V}_{(1)}^{\beta}\:\tilde{V}_{(1)\beta}}}&=&
{(1+C) \over \sqrt{(1+C)^2-D^2}}
\:{V_{(1)}^{\alpha} \over \sqrt{-V_{(1)}^{\beta}\:V_{(1)\beta}}}+
{D \over \sqrt{(1+C)^2-D^2}}
\:{V_{(2)}^{\alpha} \over \sqrt{V_{(2)}^{\beta}\:V_{(2)\beta}}}\label{2TN1}\\
{\tilde{V}_{(2)}^{\alpha}
\over \sqrt{\tilde{V}_{(2)}^{\beta}\:\tilde{V}_{(2)\beta}}}&=&
{D \over \sqrt{(1+C)^2-D^2}}
\:{V_{(1)}^{\alpha} \over \sqrt{-V_{(1)}^{\beta}\:V_{(1)\beta}}} +
{(1+C) \over \sqrt{(1+C)^2-D^2}}
\:{V_{(2)}^{\alpha} \over \sqrt{V_{(2)}^{\beta}\:V_{(2)\beta}}}\ .
\label{2TN2}
\end{eqnarray}

The condition $[(1+C)^2-D^2]>0$ enables two possible situations, $1+C > 0$ or $1+C < 0$. When $1+C > 0$, the transformations (\ref{2TN1}-\ref{2TN2})
are manifesting that an electromagnetic gauge transformation on the vector field $A^{\alpha}$, that leaves invariant the electromagnetic field $f_{\mu\nu}$,
generates a boost transformation on the normalized tetrad vector fields
$\left({V_{(1)}^{\alpha} \over \sqrt{-V_{(1)}^{\beta}\:V_{(1)\beta}}}, {V_{(2)}^{\alpha} \over \sqrt{V_{(2)}^{\beta}\:V_{(2)\beta}}}\right)$.
When the case $1+C < 0$ is fulfilled, equations (\ref{2TN1}-\ref{2TN2}) can be rewritten,

\begin{eqnarray}
{\tilde{V}_{(1)}^{\alpha}
\over \sqrt{-\tilde{V}_{(1)}^{\beta}\:\tilde{V}_{(1)\beta}}}&=&
{[-(1+C)] \over \sqrt{(1+C)^2-D^2}}
\:{\left(-V_{(1)}^{\alpha}\right)
\over \sqrt{-V_{(1)}^{\beta}\:V_{(1)\beta}}}+
{[-D] \over \sqrt{(1+C)^2-D^2}}
\:{\left(-V_{(2)}^{\alpha}\right) \over \sqrt{V_{(2)}^{\beta}
\:V_{(2)\beta}}}\label{2TN1SC}\\
{\tilde{V}_{(2)}^{\alpha}
\over \sqrt{\tilde{V}_{(2)}^{\beta}\:\tilde{V}_{(2)\beta}}}&=&
{[-D] \over \sqrt{(1+C)^2-D^2}}
\:{\left(-V_{(1)}^{\alpha}\right)
\over \sqrt{-V_{(1)}^{\beta}\:V_{(1)\beta}}} +
{[-(1+C)] \over \sqrt{(1+C)^2-D^2}}
\:{\left(-V_{(2)}^{\alpha}\right) \over \sqrt{V_{(2)}^{\beta}
\:V_{(2)\beta}}}\ . \label{2TN2SC}
\end{eqnarray}

Equations (\ref{2TN1SC}-\ref{2TN2SC}) represent the composition of two transformations. An inversion of the
normalized tetrad vector fields $\left({V_{(1)}^{\alpha} \over \sqrt{-V_{(1)}^{\beta}\:V_{(1)\beta}}},
{V_{(2)}^{\alpha} \over \sqrt{V_{(2)}^{\beta}\:V_{(2)\beta}}}\right)$, and a boost.

If the case $[(1+C)^2-D^2]<0$ is satisfied,
the vectors $V_{(1)}^{\alpha}$ and $V_{(2)}^{\alpha}$ will change their timelike or spacelike nature,

\begin{eqnarray}
\tilde{V}_{(1)}^{\alpha}\:\tilde{V}_{(1)\alpha} &=&
[-(1+C)^2+D^2]\:(-V_{(1)}^{\alpha}\:V_{(1)\alpha})\label{2FPI}\\
(-\tilde{V}_{(2)}^{\alpha}\:\tilde{V}_{(2)\alpha}) &=&
[-(1+C)^2+D^2]\:V_{(2)}^{\alpha}\:V_{(2)\alpha}\ .\label{2SPI}
\end{eqnarray}

These are special improper transformations on blade one, see references \cite{A,ROMP,SING}. The normalized tetrad vectors $V_{(1)}^{\alpha}$ and $V_{(2)}^{\alpha}$ transform as,

\begin{eqnarray}
{\tilde{V}_{(1)}^{\alpha}
\over \sqrt{\tilde{V}_{(1)}^{\beta}\:\tilde{V}_{(1)\beta}}}&=&
{(1+C) \over \sqrt{-(1+C)^2+D^2}}
\:{V_{(1)}^{\alpha} \over \sqrt{-V_{(1)}^{\beta}\:V_{(1)\beta}}}+
{D \over \sqrt{-(1+C)^2+D^2}}
\:{V_{(2)}^{\alpha} \over \sqrt{V_{(2)}^{\beta}\:V_{(2)\beta}}}\label{2TN1I}\\
{\tilde{V}_{(2)}^{\alpha}
\over \sqrt{-\tilde{V}_{(2)}^{\beta}\:\tilde{V}_{(2)\beta}}}&=&
{D \over \sqrt{-(1+C)^2+D^2}}
\:{V_{(1)}^{\alpha} \over \sqrt{-V_{(1)}^{\beta}\:V_{(1)\beta}}} +
{(1+C) \over \sqrt{-(1+C)^2+D^2}}
\:{V_{(2)}^{\alpha} \over \sqrt{V_{(2)}^{\beta}\:V_{(2)\beta}}}\ .
\label{2TN2I}
\end{eqnarray}

In manuscript \cite{A} transformations (\ref{2TN1I}-\ref{2TN2I}) have been described as ``For $D > 0$ and $1+C > 0$ these transformations
(\ref{2TN1I}-\ref{2TN2I}) represent improper space inversions on blade one. If $D > 0$ and $1+C < 0$, equations (\ref{2TN1I}-\ref{2TN2I}) are
improper time reversal transformations on blade one \cite{WE}''. This description is inaccurate, see the Erratum \cite{A} and reference \cite{SING}. Let us see why. We can rewrite transformations (\ref{2TN1I}-\ref{2TN2I}) for $D > 0$ as the composition of two different kinds of transformations. First, a local boost given by
$\Lambda^{o}_{\:\:o} = {D \over \sqrt{-(1+C)^2+D^2}}$, $\Lambda^{o}_{\:\:1} = {(1+C) \over \sqrt{-(1+C)^2+D^2}}$,
$\Lambda^{1}_{\:\:o} = {(1+C) \over \sqrt{-(1+C)^2+D^2}}$,  $\Lambda^{1}_{\:\:1} = {D \over \sqrt{-(1+C)^2+D^2}}$.
Second, a discrete transformation given by $\Lambda^{o}_{\:\:o} = 0$, $\Lambda^{o}_{\:\:1} = 1$, $\Lambda^{1}_{\:\:o} = 1$,
$\Lambda^{1}_{\:\:1} = 0$. We notice that this discrete transformation is not a Lorentz transformation because it is a discrete reflection. If the case is that $D < 0$, we can proceed
to analyze in analogy to (\ref{2TN1SC}-\ref{2TN2SC}). Then, the normalized tetrad vectors transform as,

\begin{eqnarray}
{\tilde{V}_{(1)}^{\alpha}
\over \sqrt{\tilde{V}_{(1)}^{\beta}\:\tilde{V}_{(1)\beta}}}&=&
{[-(1+C)] \over \sqrt{-(1+C)^2+D^2}}
\:{\left(-V_{(1)}^{\alpha}\right)
\over \sqrt{-V_{(1)}^{\beta}\:V_{(1)\beta}}}+
{[-D] \over \sqrt{-(1+C)^2+D^2}}
\:{\left(-V_{(2)}^{\alpha}\right) \over \sqrt{V_{(2)}^{\beta}
\:V_{(2)\beta}}}\label{2TN1SCI}\\
{\tilde{V}_{(2)}^{\alpha}
\over \sqrt{-\tilde{V}_{(2)}^{\beta}\:\tilde{V}_{(2)\beta}}}&=&
{[-D] \over \sqrt{-(1+C)^2+D^2}}
\:{\left(-V_{(1)}^{\alpha}\right)
\over \sqrt{-V_{(1)}^{\beta}\:V_{(1)\beta}}} +
{[-(1+C)] \over \sqrt{-(1+C)^2+D^2}}
\:{\left(-V_{(2)}^{\alpha}\right) \over \sqrt{V_{(2)}^{\beta}
\:V_{(2)\beta}}}\ . \label{2TN2SCI}
\end{eqnarray}

Analogously to the previous case we wrote in manuscript \cite{A} ``For $D < 0$ and $1+C < 0$ these transformations (\ref{2TN1SCI}-\ref{2TN2SCI})
represent the composition of inversions, and improper space inversions on blade one. If $D < 0$ and $1+C > 0$, equations (\ref{2TN1SCI}-\ref{2TN2SCI})
are inversions composed with improper time reversal transformations on blade one \cite{WE}''. Once again this description is inaccurate, see the Erratum \cite{A} and reference \cite{SING}. We can rewrite transformations (\ref{2TN1SCI}-\ref{2TN2SCI}) for $D < 0$ as the composition of three different kinds of transformations. First, a local boost given by $\Lambda^{o}_{\:\:o} = {[-D] \over \sqrt{-(1+C)^2+D^2}}$, $\Lambda^{o}_{\:\:1} = {[-(1+C)] \over \sqrt{-(1+C)^2+D^2}}$,
$\Lambda^{1}_{\:\:o} = {[-(1+C)] \over \sqrt{-(1+C)^2+D^2}}$,  $\Lambda^{1}_{\:\:1} = {[-D] \over \sqrt{-(1+C)^2+D^2}}$.
Second, a discrete transformation given by $\Lambda^{o}_{\:\:o} = 0$, $\Lambda^{o}_{\:\:1} = 1$, $\Lambda^{1}_{\:\:o} = 1$,  $\Lambda^{1}_{\:\:1} = 0$, which again is not a Lorentz transformation. Third, a full inversion. In plane two, the choice $Y_{\alpha} = \ast A_{\alpha} + \ast \Lambda_{,\alpha}$ induces just local spatial rotation tetrad vector transformations. For this purpose we consider equations (87-88), (89-90) and (91-92) in reference \cite{A}.

\begin{eqnarray}
M&=&(-Q/2)\:V_{(3)\sigma}\:\ast \Lambda^{\sigma} / (\:V_{(4)\beta}\:
V_{(4)}^{\beta}\:)\label{COEFFM}\\
N&=&(-Q/2)\:V_{(4)\sigma}\:\ast \Lambda^{\sigma} / (\:V_{(3)\beta}\:
V_{(3)}^{\beta}\:)\ .\label{COEFFN}
\end{eqnarray}

We would like to calculate the norm of the transformed vectors $\tilde{V}_{(3)}^{\alpha}$ and $\tilde{V}_{(4)}^{\alpha}$,

\begin{eqnarray}
\tilde{V}_{(3)}^{\alpha}\:\tilde{V}_{(3)\alpha} &=&
[(1+N)^2+M^2]\:V_{(3)}^{\alpha}\:V_{(3)\alpha}\label{FPS}\\
\tilde{V}_{(4)}^{\alpha}\:\tilde{V}_{(4)\alpha} &=&
[(1+N)^2+M^2]\:V_{(4)}^{\alpha}\:V_{(4)\alpha}\ ,\label{SPS}
\end{eqnarray}

where the relation $V_{(3)}^{\alpha}\:V_{(3)\alpha} = V_{(4)}^{\alpha}\:V_{(4)\alpha}$ has been used. We observe that the gauge transformations of the objects given by (\ref{FPS}-\ref{SPS}) cannot change the spacelike character of vectors $V_{(3)}^{\alpha}$ and $V_{(4)}^{\alpha}$, unless $1+N=M=0$. Apart from that exception, the factor $[(1+N)^2+M^2]$ is always positive, see section \ref{sec:appIII}, so we would have no problems normalizing the transformed vectors $\tilde{V}_{(3)}^{\alpha}$ and $\tilde{V}_{(4)}^{\alpha}$ for just one local electromagnetic gauge transformation $\ast \Lambda$ of the two vectors that span the local plane two,

\begin{eqnarray}
{\tilde{V}_{(3)}^{\alpha}
\over \sqrt{\tilde{V}_{(3)}^{\beta}\:\tilde{V}_{(3)\beta}}}&=&
{(1+N) \over \sqrt{(1+N)^2+M^2}}
\:{V_{(3)}^{\alpha} \over \sqrt{V_{(3)}^{\beta}\:V_{(3)\beta}}} -
{M \over \sqrt{(1+N)^2+M^2}}
\:{V_{(4)}^{\alpha} \over \sqrt{V_{(4)}^{\beta}\:V_{(4)\beta}}}\label{TN3}\\
{\tilde{V}_{(4)}^{\alpha}
\over \sqrt{\tilde{V}_{(4)}^{\beta}\:\tilde{V}_{(4)\beta}}}&=&
{M \over \sqrt{(1+N)^2+M^2}}
\:{V_{(3)}^{\alpha} \over \sqrt{V_{(3)}^{\beta}\:V_{(3)\beta}}} +
{(1+N) \over \sqrt{(1+N)^2+M^2}}
\:{V_{(4)}^{\alpha} \over \sqrt{V_{(4)}^{\beta}\:V_{(4)\beta}}}\ .
\label{TN4}
\end{eqnarray}

As long as $[(1+N)^2+M^2]>0$ the transformations (\ref{TN3}-\ref{TN4}) are saying that an electromagnetic gauge transformation with local scalar $\ast \Lambda$ on the vector field $\ast A_{\alpha} + \ast \Lambda_{\alpha}$ that leaves invariant the dual electromagnetic field $\ast f_{\mu\nu} = \ast A_{\nu ;\mu} - \ast A_{\mu ;\nu} $, generates a spatial rotation on the normalized tetrad vector fields $\left({V_{(3)}^{\alpha} \over \sqrt{V_{(3)}^{\beta}\:V_{(3)\beta}}}, {V_{(4)}^{\alpha} \over \sqrt{V_{(4)}^{\beta}\:V_{(4)\beta}}}\right)$. Let us recall that the star in $\ast A_{\nu}$ is just nomenclature, not the dual operator, meaning that $\ast A_{\nu ;\mu} = (\ast A_{\nu})_{;\mu}$. The notation for the scalar derivatives $\ast \Lambda_{,\mu} = \ast \Lambda_{\mu}$ is used for convenience and we reiterate that local tetrad electromagnetic gauge transformations can be interpreted as different new local gauge choices $X_{\alpha} = A_{\alpha} + \Lambda_{,\alpha}$ and $Y_{\alpha} = \ast A_{\alpha} + \ast \Lambda_{,\alpha}$.
For the equality $D = 1+C$ we can see using equations (\ref{2TUN}), (\ref{2TVN}) and (\ref{2E=D}-\ref{2F=C}) that,

\begin{eqnarray}
\tilde{V}_{(1)}^{\alpha} &=& (1+C)\:V_{(1)}^{\alpha} +
(1+C)\:V_{(2)}^{\alpha}\label{2TUNNULL}\\
\tilde{V}_{(2)}^{\alpha} &=& (1+C)\:V_{(2)}^{\alpha} +
(1+C)\:V_{(1)}^{\alpha}\ .\label{2TVBNULL}
\end{eqnarray}

Equations (\ref{2TUNNULL}-\ref{2TVBNULL}) show that any vector on blade one transforms as,

\begin{equation}
A\: V_{(1)}^{\alpha} + B\: V_{(2)}^{\alpha} \rightarrow
A\: \tilde{V}_{(1)}^{\alpha} + B\: \tilde{V}_{(2)}^{\alpha} = (1+C)\:(A+B)\:
( V_{(1)}^{\alpha} +  V_{(2)}^{\alpha})\ .\label{2NONINJ}
\end{equation}

This case is not trivial and a whole manuscript has been implemented to analyze it. We briefly present sections of manuscript \cite{SING} in sections \ref{timelikenull},\ref{sec:appIII} and \ref{sec:appIV} in this manuscript.

Then, after some analysis in paper \cite{A} and additionally references \cite{ROMP,SING,GLAW} the following theorems were proved,

\newtheorem {guesslb1} {Theorem}
\newtheorem {guesslb2}[guesslb1] {Theorem}
\begin{guesslb1}
The mapping between the local group of electromagnetic gauge transformations and the local group LB1 defined above is isomorphic.
\end{guesslb1}

\begin{guesslb1}
The mapping between the local group of electromagnetic gauge transformations and the local group LB2 is isomorphic. LB2 is $SO(2)$.
\end{guesslb1}

For the purpose of illustration we included section \ref{sec:appII} with the case of two composed boosts in order to prove the group law in our local tetrad electromagnetic gauge transformations. All the possible cases associated to this group mapping are provided in detail in reference \cite{GLAW}.

\section{The new tetrads determine the gravitational and the electromagnetic fields}
\label{electrounif}

Regarding the recovery of the electromagnetic potential vector from the tetrad vectors that diagonalize locally and covariantly the stress-energy tensor at every point in spacetime, we can notice the following. First let us consider the following unification tetrad version,

\begin{eqnarray}
V_{(1)unif}^{\alpha} &=& \xi^{\alpha\lambda}\:\xi_{\rho\lambda}\:X^{\rho}
\label{V1UNIF}\\
V_{(2)unif}^{\alpha} &=&  \xi^{\alpha\lambda} \: X_{\lambda}
\label{V2UNIF}\\
V_{(3)unif}^{\alpha} &=&  \ast \xi^{\alpha\lambda} \: Y_{\lambda}
\label{V3UNIF}\\
V_{(4)unif}^{\alpha} &=& \ast \xi^{\alpha\lambda}\: \ast \xi_{\rho\lambda}
\:Y^{\rho}\ .\label{V4UNIF}
\end{eqnarray}

Please notice that in vectors (\ref{V2UNIF}-\ref{V3UNIF}) there is no factor $\sqrt{-Q/2}$ as in the tetrad vectors (\ref{V2}-\ref{V3}). From the tetrad vectors (\ref{V1UNIF}-\ref{V4UNIF}) this local scalar factor will be deduced and not introduced a priori \cite{ATGU}. Let us consider the two eigenvectors of the stress-energy tensor,

\begin{eqnarray}
V_{(1)}^{\alpha} &=& \xi^{\alpha\lambda}\:\xi_{\rho\lambda}\:A^{\rho}
\label{V1U}\\
V_{(4)}^{\alpha} &=& \ast \xi^{\alpha\lambda}\: \ast \xi_{\rho\lambda}
\:A^{\rho}\ .\label{V4U}
\end{eqnarray}

Notice that the eigenvector (\ref{V4UNIF}) is gauged with $Y^{\rho}=A^{\rho}$ in (\ref{V4U}), not $\ast A^{\rho}$, see reference \cite{Addendum A}. Nonetheless, and because of its skeleton, it is inside blade two at every point in spacetime, see the whole analysis in reference \cite{A}.  Then, at the points in spacetime where the set of four vectors (\ref{V1UNIF}-\ref{V4UNIF}) is not trivial for $X^{\rho}=A^{\rho}$ and $Y^{\rho}=A^{\rho}$, we can proceed to normalize,

\begin{eqnarray}
U^{\alpha} &=& \xi^{\alpha\lambda}\:\xi_{\rho\lambda}\:A^{\rho} \:
/ \: (\: \sqrt{-Q/2} \: \sqrt{A_{\mu} \ \xi^{\mu\sigma} \
\xi_{\nu\sigma} \ A^{\nu}}\:) \label{U}\\
V^{\alpha} &=& \xi^{\alpha\lambda}\:A_{\lambda} \:
/ \: (\:\sqrt{A_{\mu} \ \xi^{\mu\sigma} \
\xi_{\nu\sigma} \ A^{\nu}}\:) \label{V}\\
Z^{\alpha} &=& \ast \xi^{\alpha\lambda} \: A_{\lambda} \:
/ \: (\:\sqrt{ A_{\mu}  \ast \xi^{\mu\sigma}
\ast \xi_{\nu\sigma}  A^{\nu}}\:)
\label{Z}\\
W^{\alpha} &=& \ast \xi^{\alpha\lambda}\: \ast \xi_{\rho\lambda}
\:A^{\rho} \: / \: (\:\sqrt{-Q/2} \: \sqrt{A_{\mu}
\ast \xi^{\mu\sigma} \ast \xi_{\nu\sigma} A^{\nu}}\:) \ .
\label{W}
\end{eqnarray}

The new expression for the metric tensor is $g_{\alpha\beta} = -U_{\alpha}\:U_{\beta} + V_{\alpha}\:V_{\beta} + Z_{\alpha}\:Z_{\beta} + W_{\alpha}\:W_{\beta}$. Using the metric tensor we can deduce the curvature of spacetime and the gravitational field. The notation we are using to name the four tetrad vectors \cite{RW,LGT8,WE,LL,HS,NP}(\ref{U}-\ref{W}) is the same notation used in \cite{HS}, even though the geometrical meaning is different. The four vectors (\ref{U}-\ref{W}) have the following algebraic properties, $-U^{\alpha}\:U_{\alpha}=V^{\alpha}\:V_{\alpha}=Z^{\alpha}\:Z_{\alpha}=W^{\alpha}\:W_{\alpha}=1$. Any other scalar product is zero. The point is that taking the difference between both vectors (\ref{V1U}-\ref{V4U}) and using the identity (\ref{2i3}) we get,

\begin{eqnarray}
V_{(1)}^{\alpha} - V_{(4)}^{\alpha} = \xi^{\alpha\lambda}\:\xi_{\rho\lambda}\:A^{\rho} - \ast \xi^{\alpha\lambda}\: \ast \xi_{\rho\lambda}\:A^{\rho} = \frac{1}{2}\: Q \: A^{\alpha} .\label{recovery}
\end{eqnarray}

Next we calculate the following scalar products, $V_{(1)}^{\alpha}\:V_{(1)\alpha}=(-Q/2) \: (A_{\mu} \ \xi^{\mu\sigma} \
\xi_{\nu\sigma} \ A^{\nu}\:)=(-Q/2)\:V_{(2)}^{\alpha}\:V_{(2)\alpha}$. It is also easy to verify that the following scalar products hold , $V_{(4)}^{\alpha}\:V_{(4)\alpha}=(-Q/2) \: (A_{\mu} \ \ast \xi^{\mu\sigma} \
\ast \xi_{\nu\sigma} \ A^{\nu}\:)=(-Q/2)\:V_{(3)}^{\alpha}\:V_{(3)\alpha}$. Then we call $C_{aux} = (V_{(1)}^{\alpha}\:V_{(1)\alpha})/(V_{(2)}^{\alpha}\:V_{(2)\alpha})=(V_{(4)}^{\alpha}\:V_{(4)\alpha})/(V_{(3)}^{\alpha}\:V_{(3)\alpha})=(-Q/2)$. It is elementary to realize that the electromagnetic vector potential can be expressed as $A^{\alpha}=-(V_{(1)}^{\alpha} - V_{(4)}^{\alpha})/C_{aux}$. That is, we can express the electromagnetic potential just using the tetrad vectors (\ref{V1UNIF}-\ref{V4UNIF}) for the choice $X^{\rho}=A^{\rho}$ and $Y^{\rho}=A^{\rho}$. Evidently from the potential we will be able to find the electromagnetic field $f_{\mu\nu}$ and using the techniques of section \ref{intro} we would be able to find the extremal field $\xi_{\mu\nu}$, the local gauge invariant $Q=\xi_{\mu\nu}\:\xi^{\mu\nu}$ and the complexion local gauge invariant $\alpha$ through $\tan(2\alpha) = - f_{\mu\nu}\:\ast f^{\mu\nu} / f_{\lambda\rho}\:f^{\lambda\rho}$. The extremal field tensor and its dual can then be written, $\xi_{\alpha\beta} = -2\:\sqrt{-Q/2}\:U_{[\alpha}\:V_{\beta]}$ and $\ast \xi_{\alpha\beta} = 2\:\sqrt{-Q/2}\:Z_{[\alpha}\:W_{\beta]}$. These expressions for the extremal field and its dual are providing the necessary information to express the electromagnetic field in terms of the new tetrad,
$f_{\alpha\beta} = -2\:\sqrt{-Q/2}\:\:\cos\alpha\:\:U_{[\alpha}\:V_{\beta]} + 2\:\sqrt{-Q/2}\:\:\sin\alpha\:\:Z_{[\alpha}\:W_{\beta]}$, see reference \cite{A} for all the details. These new tetrads are grand unification objects.

\section{Timelike and spacelike vectors transform into null vectors via electromagnetic gauge transformations}
\label{timelikenull}

Even though in our original paper \cite{A} we are dealing with vacuum Maxwell equations without source terms we will proceed to analyze the Coulomb case, which shares similarities in gauge analysis with the Reissner-Nordstr\"{o}m case which is a solution to the vacuum Einstein-Maxwell equations, for instance on the local plane one.
The results in section \ref{localbladeone} are also valid for the Maxwell equations with sources $J^{\mu}$ in Minkowski spacetime were the particular tetrad construction and gauge analysis are presented in section \ref{sec:appV}.

%We remind ourselves that in the submitted manuscript we also work with vacuum Maxwell equations without sources on a flat Minkowski background which allowed for the existence of two potentials $X^{\alpha}=A^{\alpha}$ and $Y^{\alpha}=\ast A^{\alpha}$, see reference \cite{A} and sections \ref{potentials} and \ref{gaugegeometry}.

Let us start \cite{SING} with the case where the gauge choice is $f_{tr} = e/r^{2}$, $A_{t} = e/r$ and $A_{r} = 0$. Let us analyze the components of the tetrad vectors (\ref{V1}-\ref{V2}) for this case for a flat Minkowskian spacetime with signature $(-+++)$ when the choice is $X^{\rho}=A^{\rho}$.

\begin{eqnarray}
V_{(1)}^{t} &=& \xi^{tr}\:\xi_{tr}\:A^{t} = \mid \xi_{tr} \mid^{2}\:A_{t} \label{V1tEX}\\
V_{(1)}^{r} &=& \xi^{rt}\:\xi_{rt}\:A^{r} = 0 \label{V1rEX}\\
V_{(2)}^{t} &=& \mid \xi_{tr} \mid \: \xi^{tr} \: A_{r} = 0 \label{V2tEX}\\
V_{(2)}^{r} &=& \mid \xi_{tr} \mid \: \xi^{rt} \: A_{t} = \mid \xi_{tr} \mid\:\xi_{tr}\:A_{t} \ .\label{V2rEX}
\end{eqnarray}

where $Q = -2\mid \xi_{tr} \mid^{2}$. Next let us proceed to analyze the norm of these different orthogonal vectors.

\begin{eqnarray}
\lefteqn{ V_{(1)}^{\alpha}\:V_{(1)\alpha} = V_{(1)}^{t}\:V_{(1)t} + V_{(1)}^{r}\:V_{(1)r} }\nonumber \\ &&= -\mid \xi_{tr} \mid^{4}\:(A_{t})^{2} +
\mid \xi_{tr} \mid^{4}\:(A_{r})^{2} = -\mid \xi_{tr} \mid^{4}\:(A_{t})^{2} \label{FPCB}\\
&& V_{(2)}^{\alpha}\:V_{(2)\alpha} = V_{(2)}^{t}\:V_{(2)t} + V_{(2)}^{r}\:V_{(2)r}  \nonumber \\ &&= -\mid \xi_{tr} \mid^{4}\:(A_{r})^{2} +
\mid \xi_{tr} \mid^{4}\:(A_{t})^{2} = \mid \xi_{tr} \mid^{4}\:(A_{t})^{2} \ .\label{SPCB}
\end{eqnarray}

where the relation $V_{(1)}^{\alpha}\:V_{(1)\alpha} = -V_{(2)}^{\alpha}\:V_{(2)\alpha} \ne 0$ is evident.

We then proceed immediately to the following particular case. The Coulomb example where $f_{tr} = e/r^{2} = \xi_{tr}$, $A_{t} = e/r$ and $A^{new}_{r} = -e/r$. If we do not write ```new'' we mean the original components before the local electromagnetic gauge transformation. We reiterate that all the results in section \ref{localbladeone} are also valid for the Maxwell equations with sources $J^{\mu}$ in Minkowski spacetime, please see section \ref{sec:appV}. The analysis in section \ref{localbladeone} is also valid for Einstein-Maxwell equations in curved spacetimes with sources $J^{\mu}$. Resuming our discussion this Coulomb case corresponds to a gauge transformation $\Lambda_{t} = 0$ and $\Lambda_{r} = -e/r$ of the original gauge choice $A_{t} = e/r$ and $A_{r} = 0$. We notice that a local electromagnetic gauge transformation of the ``gauge vectors'' $X^{\alpha}=A^{\alpha}$ and $Y^{\alpha}$ can be just interpreted as a new choice for the gauge vectors $X_{\alpha} = A_{\alpha} + \Lambda_{,\alpha}$ and $Y_{\alpha} \rightarrow Y_{\alpha} + \ast \Lambda_{,\alpha}$. We notice that we are not specifying or choosing the gauge vector $Y^{\alpha}$ in the case with sources $J^{\mu}$ because we leave this choice for the discussion in section \ref{sec:appV}. For simplicity we will use the notation for local gauge transformations $\Lambda_{,\mu} = \Lambda_{\mu}$ where $\Lambda$ is a local scalar. Let us analyze the components of the tetrad vectors (\ref{V1}-\ref{V2}) for the new case raised for a flat Minkowskian spacetime with signature $(-+++)$.

\begin{eqnarray}
\tilde{V}_{(1)}^{t} &=& \xi^{tr}\:\xi_{tr}\:A^{t} = \mid \xi_{tr} \mid^{2}\:A_{t} \label{V1tEX}\\
\tilde{V}_{(1)}^{r} &=& \xi^{rt}\:\xi_{rt}\:A_{new}^{r} = -\mid \xi_{tr} \mid^{2}\:A^{new}_{r} \label{V1rEX}\\
\tilde{V}_{(2)}^{t} &=& \mid \xi_{tr} \mid \: \xi^{tr} \: A^{new}_{r} = -\mid \xi_{tr} \mid\:\xi_{tr}\:A^{new}_{r} \label{V2tEX}\\
\tilde{V}_{(2)}^{r} &=& \mid \xi_{tr} \mid \: \xi^{rt} \: A_{t} = \mid \xi_{tr} \mid\:\xi_{tr}\:A_{t} \ .\label{V2rEX}
\end{eqnarray}

where $Q = -2\mid \xi_{tr} \mid^{2}$. Next let us proceed to analyze the norm of these vectors.

\begin{eqnarray}
\lefteqn{ \tilde{V}_{(1)}^{\alpha}\:\tilde{V}_{(1)\alpha} = \tilde{V}_{(1)}^{t}\:\tilde{V}_{(1)t} + \tilde{V}_{(1)}^{r}\:\tilde{V}_{(1)r} }\nonumber \\ &&= -\mid \xi_{tr} \mid^{4}\:(A_{t})^{2} +
\mid \xi_{tr} \mid^{4}\:(A^{new}_{r})^{2} = \mid \xi_{tr} \mid^{4}\:(-(A_{t})^{2} + (A^{new}_{r})^{2})=0 \label{FPC}\\
&& \tilde{V}_{(2)}^{\alpha}\:\tilde{V}_{(2)\alpha} = \tilde{V}_{(2)}^{t}\:\tilde{V}_{(2)t} + \tilde{V}_{(2)}^{r}\:\tilde{V}_{(2)r}  \nonumber \\ &&= -\mid \xi_{tr} \mid^{4}\:(A^{new}_{r})^{2} +
\mid \xi_{tr} \mid^{4}\:(A_{t})^{2} = \mid \xi_{tr} \mid^{4}\:((A_{t})^{2} - (A^{new}_{r})^{2})=0 \ .\label{SPC}
\end{eqnarray}

where the relation $\tilde{V}_{(1)}^{\alpha}\:\tilde{V}_{(1)\alpha} = -\tilde{V}_{(2)}^{\alpha}\:\tilde{V}_{(2)\alpha}$ is evident. From the detailed analysis reproduced from reference \cite{A} in section IV ``gauge geometry'' we calculate the coefficients C and D from equations (54-55) in that manuscript also provided in (\ref{2COEFFC}-\ref{2COEFFD}). We then will identify the coefficients in equations (\ref{FPC}-\ref{SPC}) for the Coulomb case with the coefficients in the general equations (56-57) from reference \cite{A} given by $\tilde{V}_{(1)}^{\alpha}\:\tilde{V}_{(1)\alpha} = [(1+C)^2-D^2]\:V_{(1)}^{\alpha}\:V_{(1)\alpha}$ and $\tilde{V}_{(2)}^{\alpha}\:\tilde{V}_{(2)\alpha} =
[(1+C)^2-D^2]\:V_{(2)}^{\alpha}\:V_{(2)\alpha}$.

\begin{eqnarray}
C&=&(-Q/2)\:V_{(1)\sigma}\:\Lambda^{\sigma} / (\:V_{(2)\beta}\:
V_{(2)}^{\beta}\:) = \mid \xi_{tr} \mid^{2}\:{(\Lambda^{t}\:V_{(1)t} + \Lambda^{r}\:V_{(1)r}) \over \mid \xi_{tr} \mid^{4}\:((A_{t})^{2} + (A_{r})^{2})}\label{COEFFCAT1}\\
D&=&(-Q/2)\:V_{(2)\sigma}\:\Lambda^{\sigma} / (\:V_{(1)\beta}\:
V_{(1)}^{\beta}\:) = \mid \xi_{tr} \mid^{2}\:{(\Lambda^{t}\:V_{(2)t} + \Lambda^{r}\:V_{(2)r}) \over \mid \xi_{tr} \mid^{4}\:(-(A_{t})^{2} + (A_{r})^{2})}\ .\label{COEFFDAT2}
\end{eqnarray}

Both these coefficients (\ref{COEFFCAT1}-\ref{COEFFDAT2}) after some simple algebra reduce to,

\begin{eqnarray}
C&=& -{(\Lambda^{t}\:A_{t} + \Lambda^{r}\:A_{r}) \over ((A_{t})^{2} + (A_{r})^{2})}\label{COEFFCBT1}\\
D&=& {\xi_{tr} \over \mid \xi_{tr} \mid}\:{(\Lambda^{t}\:A_{r} + \Lambda^{r}\:A_{t}) \over (-(A_{t})^{2} + (A_{r})^{2})}\ .\label{COEFFDBT2}
\end{eqnarray}

In the last general equation (\ref{COEFFCBT1}) we notice the following. The original gauge from which we transform is $A_{t} = e/r$ and $A_{r} = 0$. The local gauge transformation is $\Lambda_{t} = 0$ and $\Lambda_{r} = -e/r$. Therefore, the $A_{r}$ in the coefficient C is zero, because it is the old $A_{r}$. Therefore, $C = 0$. For the second equation (\ref{COEFFDBT2}) knowing that $\xi_{tr} = f_{tr} = e/r^{2}$, we find $D =-{\Lambda^{r} \over A_{t}} = -{\Lambda_{r} \over A_{t}}$.
Finally, we calculate the norm transformation coefficient in equations (56-57) of manuscript \cite{A}.

\begin{eqnarray}
[(1+C)^2-D^2] = (1-0)^{2} - ({\Lambda_{r} \over A_{t}})^{2} = {A_{t}^{2} - \Lambda_{r}^{2} \over A_{t}^{2}} = 0 \label{NULLCASE}
\end{eqnarray}

The same result can be noticed from the particular analysis in equations (\ref{FPC}-\ref{SPC}). Because the new gauge is a very special gauge for which $A_{t} = - \Lambda_{r} = e/r$. We will prove in the next section that it is the inhomogeneous solution to a differential equation. For this particular gauge transformation both original vectors (\ref{V1}-\ref{V2}) one timelike and the other spacelike are transformed into null vectors on the local light cone. It is a singular and unique gauge transformation. Only one in an infinite set. It is in fact a set of measure zero in the whole set of gauge transformations.

\subsection{Gauge differential equation: Coulomb case}
\label{diffeq}

All the results in section \ref{localbladeone} are also valid for the Maxwell equations with sources $J^{\mu}$ in Minkowski spacetime were the particular tetrad construction and gauge analysis are presented in section \ref{sec:appV}. In this section we will proceed following an inverse path. We will impose the null condition for gauge vector transformation and from the ensuing differential equation on the local scalar we will find the local gauge transformations that take a timelike and a spacelike vectors into the same null vector on the local light cone. Let us then impose the condition $D = 1+ C$ in accordance to the general theory of tetrad gauge transformations through equations (56-57) in reference \cite{A}. We briefly remind ourselves that two cases are possible for $[(1+C)^2-D^2]=0$ and the case $D = -(1+ C)$ is analyzed in a similar way leading to similar results. Next we transform from $A_{t} = e/r$ and $A_{r} = 0$. We will not include hereafter the ```new'' label because we do not want to overload with notation and because it is easy to follow the components before and after the local gauge transformations. The general local gauge transformation is in principle $\Lambda_{t} \ne 0$ and $\Lambda_{r} \ne 0$. Let us use the general equations (\ref{COEFFCBT1}-\ref{COEFFDBT2}) for the coefficients C and D in the equation $D = 1+ C$,

\begin{eqnarray}
{\xi_{tr} \over \mid \xi_{tr} \mid}\:{(\Lambda^{t}\:A_{r} + \Lambda^{r}\:A_{t}) \over (-(A_{t})^{2} + (A_{r})^{2})} = 1 + (-){(\Lambda^{t}\:A_{t} + \Lambda^{r}\:A_{r}) \over ((A_{t})^{2} + (A_{r})^{2})} \label{generalD+C}
\end{eqnarray}

We start from $A_{t} = e/r$ and $A_{r} = 0$ and $\xi_{tr} = f_{tr} = e/r^{2}$. Therefore, we are left with,

\begin{eqnarray}
-{\Lambda^{r} \over A_{t}} = 1 - {\Lambda^{t} \over A_{t}} \label{generalD+Cparticular}
\end{eqnarray}

We can rewrite this equation as,

\begin{eqnarray}
\Lambda^{t} - \Lambda^{r} = A_{t} \ , \label{gaugediffeq}
\end{eqnarray}

which is equivalent to,

\begin{eqnarray}
-\Lambda_{t} - \Lambda_{r} = A_{t} \ , \label{gaugediffeq2}
\end{eqnarray}

Taking cross derivatives with respect to coordinates $t$ and $r$ and reminding about the integrability condition $\Lambda_{tr} = \Lambda_{rt}$ and after some simple algebra knowing that $\partial_{t}A_{t} = 0$ we find,

\begin{eqnarray}
\Lambda_{tt} - \Lambda_{rr} = \partial_{r}A_{t} \ . \label{gaugediffeq3}
\end{eqnarray}

The inhomogeneous solution is $\Lambda_{r} = -e/r = -A_{t}$. The homogeneous solution can be found to be $\Lambda_{H} = A\:\cos\omega(t-r) + B\:\sin\omega(t-r)$ which are gauge waves traveling to the future at the speed of light where $A$, $B$ and $\omega$ are constants. We are contemplating not only equation (\ref{gaugediffeq3}) but also (\ref{gaugediffeq2}). It is just one inhomogeneous gauge transformation, therefore a set of measure zero. The inhomogeneous solution for $D=-(1+C)$ in the past light cone will correspond to $\Lambda_{r} = e/r = A_{t}$. Reflections about the light cone leave the solutions already found invariant.

\subsection{Gauge differential equation: Reissner-Nordstr\"{o}m case}
\label{diffeqrn}

The line element for this spacetime is given by the following expression \cite{RW,LGT8,AMONO},

\begin{eqnarray}
ds^{2} = - (1 - {2m \over r} + {e^{2} \over r^{2}})\: dt^{2} + (1 - {2m \over r} + {e^{2} \over r^{2}})^{-1}\: dr^{2} + r^{2}\:(d\theta^{2} + \sin^{2}\theta\:d\phi^{2})\ . \label{reissnord}
\end{eqnarray}

In this section we will proceed following the same path as in section \ref{diffeq} for the Coulomb case. Once again we will impose the null condition for gauge vector transformation and from the ensuing differential equation on the local scalar we will find the local gauge transformations that take a timelike and a spacelike vectors into the same null vector on the local light cone \cite{SING}. This time on a solution to the vacuum Einstein-Maxwell equations without sources in the Maxwell equations. Let us then impose the condition $D = 1+ C$ in accordance to the general theory of tetrad gauge transformations through equations (56-57) in reference \cite{A}. Next we transform from $A_{t} = e/r$ and $A_{r} = 0$. The general local gauge transformation is in principle $\Lambda_{t} \ne 0$ and $\Lambda_{r} \ne 0$. Let us analyze the components of the tetrad vectors (\ref{V1}-\ref{V2}) for the Reissner-Nordstr\"{o}m case with signature $(-+++)$.

\begin{eqnarray}
V_{(1)}^{t} &=& \xi^{tr}\:\xi_{tr}\:A^{t} = -g^{tt}\mid \xi_{tr} \mid^{2}\:A_{t} \label{V1tRN}\\
V_{(1)}^{r} &=& \xi^{rt}\:\xi_{rt}\:A^{r} = -g^{rr}\mid \xi_{tr} \mid^{2}\:A_{r} \label{V1rRN}\\
V_{(2)}^{t} &=& \mid \xi_{tr} \mid \: \xi^{tr} \: A_{r} = -\mid \xi_{tr} \mid\:\xi_{tr}\:A_{r} \label{V2tRN}\\
V_{(2)}^{r} &=& \mid \xi_{tr} \mid \: \xi^{rt} \: A_{t} = \mid \xi_{tr} \mid\:\xi_{tr}\:A_{t} \ .\label{V2rRN }
\end{eqnarray}

where $Q = -2\mid \xi_{tr} \mid^{2}$. Next let us proceed to analyze the norm of these vectors.

\begin{eqnarray}
\lefteqn{ V_{(1)}^{\alpha}\:V_{(2)\alpha} = V_{(1)}^{t}\:V_{(2)t} + V_{(1)}^{r}\:V_{(2)r} }\nonumber \\ &&= g^{tt}\:g_{tt}\:\mid \xi_{tr} \mid^{3}\:\xi_{tr}\:A_{t}\:A_{r} -
g^{rr}\:g_{rr}\:\mid \xi_{tr} \mid^{3}\:\xi_{tr}\:A_{r}\:A_{t} \nonumber \\ &&= \mid \xi_{tr} \mid^{3}\:\xi_{tr}\:A_{t}\:A_{r}\:(g^{tt}\:g_{tt}-g^{rr}\:g_{rr}) = 0 \label{FSPCRN}\\
\lefteqn{ V_{(1)}^{\alpha}\:V_{(1)\alpha} = V_{(1)}^{t}\:V_{(1)t} + V_{(1)}^{r}\:V_{(1)r} }\nonumber \\ &&= \mid \xi_{tr} \mid^{4}\:(g^{tt}\:(A_{t})^{2} +
g^{rr}\:(A_{r})^{2}) = -\mid \xi_{tr} \mid^{4}\:(g_{rr}\:(A_{t})^{2} + g_{tt}\:(A_{r})^{2}) \label{FPCRN}\\
&& V_{(2)}^{\alpha}\:V_{(2)\alpha} = V_{(2)}^{t}\:V_{(2)t} + V_{(2)}^{r}\:V_{(2)r} = \mid \xi_{tr} \mid^{4}\:(g_{rr}\:(A_{t})^{2} + g_{tt}\:(A_{r})^{2}) \ .\label{SPCRN}
\end{eqnarray}

where the relation $V_{(1)}^{\alpha}\:V_{(1)\alpha} = -V_{(2)}^{\alpha}\:V_{(2)\alpha}$ is evident. From the detailed analysis reproduced from reference \cite{A} in section IV ``gauge geometry'' we calculate the coefficients C and D from equations (54-55) in that manuscript. We then will identify the coefficients in equations (\ref{FPCRN}-\ref{SPCRN}) for the Reissner-Nordstr\"{o}m case with the coefficients in the general equations (56-57) in reference \cite{A} $\tilde{V}_{(1)}^{\alpha}\:\tilde{V}_{(1)\alpha} = [(1+C)^2-D^2]\:V_{(1)}^{\alpha}\:V_{(1)\alpha}$ and $\tilde{V}_{(2)}^{\alpha}\:\tilde{V}_{(2)\alpha} = [(1+C)^2-D^2]\:V_{(2)}^{\alpha}\:V_{(2)\alpha}$.

\begin{eqnarray}
C&=&(-Q/2)\:V_{(1)\sigma}\:\Lambda^{\sigma} / (\:V_{(2)\beta}\:
V_{(2)}^{\beta}\:) = \mid \xi_{tr} \mid^{2}\:{(\Lambda^{t}\:V_{(1)t} + \Lambda^{r}\:V_{(1)r}) \over \mid \xi_{tr} \mid^{4}\:(g_{rr}\:(A_{t})^{2} + g_{tt}\:(A_{r})^{2})}\label{COEFFCAT1RN}\\
D&=&(-Q/2)\:V_{(2)\sigma}\:\Lambda^{\sigma} / (\:V_{(1)\beta}\:
V_{(1)}^{\beta}\:) = -\mid \xi_{tr} \mid^{2}\:{(\Lambda^{t}\:V_{(2)t} + \Lambda^{r}\:V_{(2)r}) \over \mid \xi_{tr} \mid^{4}\:(g_{rr}\:(A_{t})^{2} + g_{tt}\:(A_{r})^{2})}\ .\label{COEFFDAT2RN}
\end{eqnarray}

Both these coefficients (\ref{COEFFCAT1RN}-\ref{COEFFDAT2RN}) after some simple algebra reduce to,

\begin{eqnarray}
C&=&  -{(\Lambda^{t}\:A_{t} + \Lambda^{r}\:A_{r}) \over (g_{rr}\:(A_{t})^{2} + g_{tt}\:(A_{r})^{2})}\label{COEFFCBT1RN}\\
D&=&  -{\xi_{tr} \over \mid \xi_{tr} \mid}\:{(-g_{tt}\:\Lambda^{t}\:A_{r} + g_{rr}\:\Lambda^{r}\:A_{t}) \over (g_{rr}\:(A_{t})^{2} + g_{tt}\:(A_{r})^{2})}\ .\label{COEFFDBT2RN}
\end{eqnarray}

Let us then impose the singular condition $D = 1+C$,

\begin{eqnarray}
-{\xi_{tr} \over \mid \xi_{tr} \mid}\:{(-g_{tt}\:\Lambda^{t}\:A_{r} + g_{rr}\:\Lambda^{r}\:A_{t}) \over (g_{rr}\:(A_{t})^{2} + g_{tt}\:(A_{r})^{2})}\  = 1 + (-){(\Lambda^{t}\:A_{t} + \Lambda^{r}\:A_{r}) \over (g_{rr}\:(A_{t})^{2} + g_{tt}\:(A_{r})^{2})}\label{generalD+CRN}
\end{eqnarray}

Let us consider the original gauge as $A_{t} = e/r$ and $A_{r} = 0$ and $\xi_{tr} = f_{tr} = e/r^{2}$, knowing that $\Lambda^{t} = g^{tt}\:\Lambda_{t}$ and $\Lambda^{r} = g^{rr}\:\Lambda_{r}$. Therefore, we are left with,

\begin{eqnarray}
{g_{tt}\:\Lambda_{r} \over A_{t}} = 1 + {\Lambda_{t} \over A_{t}} \label{generalD+CparticularRN}
\end{eqnarray}

We can rewrite this equation as,

\begin{eqnarray}
\Lambda_{t} - g_{tt}\:\Lambda_{r} = -A_{t} \ , \label{gaugediffeqRN}
\end{eqnarray}

%which is equivalent to,

%\begin{eqnarray}
%-\Lambda_{t} - \Lambda_{r} = A_{t} \ , \label{gaugediffeq2}
%\end{eqnarray}

Taking cross derivatives with respect to coordinates $t$ and $r$ and reminding about the integrability condition $\Lambda_{tr} = \Lambda_{rt}$ and after some simple algebra knowing that $\partial_{t}A_{t} = 0$ we find,

\begin{eqnarray}
\Lambda_{tt} - g_{tt}\:\partial_{r}g_{tt} \Lambda_{r} - (g_{tt})^{2} \Lambda_{rr}= -g_{tt}\:\partial_{r}A_{t} \ . \label{gaugediffeq3RN}
\end{eqnarray}

The inhomogeneous solution is $\Lambda_{r} = -g_{rr}\:A_{t}$. The homogeneous solution can be found to be a linear combination of the real and imaginary parts of,

\begin{eqnarray}
\Lambda_{H} = \exp(-\imath\int_{a}^{r}({\omega \over g_{tt}} + k)\:dr)\:\exp\imath(k\:r-\omega\:t) \ , \label{INHRN}
\end{eqnarray}

which are gauge waves with $a$, $k$ and $\omega$ constants. We are contemplating not only equation (\ref{gaugediffeq3RN}) but also (\ref{gaugediffeqRN}). There is only a unique inhomogeneous solution $\Lambda_{r} = -{e \over r\:(1 - {2m \over r} + {e^{2} \over r^{2}})} = A_{t}/g_{tt}$. It is just one inhomogeneous gauge transformation, therefore a set of measure zero. The inhomogeneous solution for $D=-(1+C)$ in the past light cone will correspond to $\Lambda_{r} = g_{rr}\:A_{t}$. Reflections about the light cone leave the solutions already found invariant.

\subsection{General differential equation}
\label{geneq}

In this final section we will present the general differential equation from which all the particular problems previously studied arise \cite{SING}.

\begin{eqnarray}
C&=&(-Q/2)\:V_{(1)\sigma}\:\Lambda^{\sigma} / (\:V_{(2)\beta}\:
V_{(2)}^{\beta}\:) \label{COEFFCGEN}\\
D&=&(-Q/2)\:V_{(2)\sigma}\:\Lambda^{\sigma} / (\:V_{(1)\beta}\:
V_{(1)}^{\beta}\:) \ .\label{COEFFDGEN}
\end{eqnarray}

where the relation $V_{(1)}^{\alpha}\:V_{(1)\alpha} = -V_{(2)}^{\alpha}\:V_{(2)\alpha}$ has been used. When we impose the condition $D = 1+ C$ we obtain the following differential equation on the local scalar gradient $\Lambda^{\sigma}$.

\begin{eqnarray}
(-Q/2)\:V_{(2)\sigma}\:\Lambda^{\sigma} / (\:V_{(1)\beta}\:
V_{(1)}^{\beta}\:) = 1 + (-Q/2)\:V_{(1)\sigma}\:\Lambda^{\sigma} / (\:V_{(2)\beta}\:
V_{(2)}^{\beta}\:) \ . \label{DIFFEQGEN}
\end{eqnarray}

By multiplying both sides by $V_{(1)}^{\alpha}\:V_{(1)\alpha} = -V_{(2)}^{\alpha}\:V_{(2)\alpha}$ we get,

\begin{eqnarray}
(-Q/2)\:V_{(2)\sigma}\:\Lambda^{\sigma}  =
V_{(1)\beta}\:V_{(1)}^{\beta} - (-Q/2)\:V_{(1)\sigma}\:\Lambda^{\sigma}  \ . \label{DIFFEQGEN1}
\end{eqnarray}

This is a differential equation on the gradient $\Lambda^{\sigma}$ with a source term $V_{(1)\beta}\:V_{(1)}^{\beta}$. Therefore it will possess an inhomogeneous solution and homogeneous solutions. Let us not forget that for the case $D = -(1+C)$ there is also one more singular inhomogeneous solution causing the timelike and spacelike vectors on the local plane one to transform into null vectors on the local light cone. This additional solution $D = -(1+C)$ corresponds to an inhomogeneous solution in the past light cone. Reflections about the light cone leave the solutions already found invariant.

\section{New physical prediction in geometrodynamics}
\label{newphysical}

Let us consider that we start with a local electromagnetic gauge transformation $\Lambda$ that induces or generates a boost on the two vectors that span the local plane one as proven in detail in manuscript \cite{A} and section \ref{intro}. It has been subsequently proven that when we multiply $\Lambda$ by a suitable real constant factor $n$, the resulting local scalar $n\:\Lambda$ generates a local electromagnetic gauge transformation that inverts the future directed timelike vectors into past directed timelike vectors in a region of spacetime inside a null surface, see reference \cite{A}. There is a local scalar factor between the norm of the original vectors that span the local plane one and the transformed vectors under the corresponding local gauge transformation. This factor has been expressed as $[(1+C)^2-D^2]$ and it is of a kinematic nature. We called the vectors that span the plane one $V_{(1)}^{\alpha}$ and $V_{(2)}^{\alpha}$. Both are eigenvectors of the stress-energy tensor with the same eigenvalue $Q/2$, see references \cite{A,ATGU} for all the details and we also define \cite{A} the local scalars $C$ and $D$ as (see also section \ref{intro})

\begin{eqnarray}
C&=&(-Q/2)\:V_{(1)\sigma}\:\Lambda^{\sigma} / (\:V_{(2)\beta}\:
V_{(2)}^{\beta}\:)\label{COEFFC}\\
D&=&(-Q/2)\:V_{(2)\sigma}\:\Lambda^{\sigma} / (\:V_{(1)\beta}\:
V_{(1)}^{\beta}\:)\ ,\label{COEFFD}
\end{eqnarray}

where $Q = -\sqrt{T_{\mu\nu}T^{\mu\nu}}=\xi_{\mu\nu}\:\xi^{\mu\nu}$ and $T^{\mu\nu}$ is the Einstein-Maxwell stress-energy tensor according to equations (39) in \cite{MW}. $Q$ as said before is assumed not to be zero, because we are dealing with non-null electromagnetic fields. If the case is that $[(1+C)^2-D^2]>0$, the vectors $V_{(1)}^{\alpha}$ and $V_{(2)}^{\alpha}$ will not change their timelike or spacelike nature or character,

\begin{eqnarray}
(-\tilde{V}_{(1)}^{\alpha}\:\tilde{V}_{(1)\alpha}) &=&
[(1+C)^2-D^2]\:(-V_{(1)}^{\alpha}\:V_{(1)\alpha})\label{FPI}\\
\tilde{V}_{(2)}^{\alpha}\:\tilde{V}_{(2)\alpha} &=&
[(1+C)^2-D^2]\:V_{(2)}^{\alpha}\:V_{(2)\alpha}\ .\label{SPI}
\end{eqnarray}

These are proper transformations on blade one. By performing a local gauge transformation by a local scalar $n\:\Lambda$ in a region of spacetime inside a null surface, we can change timelike future oriented vectors with $1+C>0$ into timelike past oriented vectors with $1+C<0$ in the local plane one, see equations (\ref{2TN1SC}-\ref{2TN2SC}). Equivalently we can also say that we manage as we will discuss in section \ref{ahatimeinv} to turn a coefficient $1+C>0$ into a coefficient of the kind $1+C<0$. The reason is that these structures arise in Einstein-Maxwell spacetimes where the notion of electromagnetic gauge is available. The fundamental point that we are making is that we can obtain this result with local Abelian gauge transformations which is a new and fundamental finding. By allowing an electromagnetic local gauge transformation in the potentials we can change the local nature of spacetime in a region where a suitable $n\:\Lambda$ has this result or effect. Electromagnetic phenomena has the causality implication \cite{RW,LL,HS,NP,CBDW,CW} that enables a jump inside the region of validity from $[(1+C)^2-D^2]>0$ and $1+C>0$ into $[(1+C)^2-D^2]>0$ and $1+C<0$, see reference \cite{FULL}. This is a new and important development in relativity.

\subsection{Aharonov-Bohm setup for the full inversion experiment}
\label{ahatimeinv}

The Aharonov-Bohm effect arises in quantum mechanics when the inclusion of a potential results in the introduction of a phase in the wave function of the electron. This phase has no consequence on the observed behavior of the electron because when a property is measured the amplitude of the wave function is involved and not its phase. However, this phase can be detected when measuring the quantum mechanical interference between electrons that have taken two different paths from a source to a detector. If these paths travel through regions with different local values of gauge potential then a difference in phase will change the measured interference pattern. This effect was found by theoretical means in 1959 by Yakir Aharonov and David Bohm and confirmed by an experiment carried out by Robert Chambers in 1960. Chambers sent electrons on different paths that passed next to a very long solenoid. The magnetic field outside such a solenoid was negligible however the vector potential outside the solenoid was significant and local. In the end electrons taking the different paths around the solenoid acquire different phases, see references \cite{AB1,AB2,S,Y}. In this section we will be dedicated to test a theoretical-experimental model for inverting the vectors in the local lightcone, see reference \cite{SCR} for the case of a spacetime reflection. To this end we will hypothesize to build a solenoid in the Aharonov-Bohm spirit \cite{AB1,AB2,S} with the purpose of having no electromagnetic fields outside even though the electromagnetic four-potential will not be zero, see for example reference \cite{BF}. There is a curl-free vector potential outside the solenoid with non-trivial enclosed flux. Let us also imagine that in the exterior of the solenoid \cite{NASA} we managed to create a constant magnetic field pointing in the direction of the solenoid axis. On one hand the solenoid will play the role of an Aharonov-Bohm topological element making spacetime not-simply connected. On the other hand the exterior magnetic field will make an electron move in circles of radius $R_{e}$ outside and around the solenoid such that $R_{disk} > R_{e} > R_{sol}$. $R_{sol}$ is the solenoid radius,  $R_{disk}$ will be the radius of the exterior disk where we have set up a constant possibly different magnetic field pointing parallel to the solenoid symmetry axis. The current in the solenoid coil will create a magnetic flux that we will call $\Phi_{sol}$. Because of the non-trivial Aharonov-Bohm topology, the electron Dirac wavefunction $\Psi$ will acquire a non-trivial phase every time the electron completes a circle around the solenoid. The phase will be given by an exponential of the expression $\frac{e}{\hbar}\:\Lambda = \frac{e}{\hbar}\:\{\Phi_{sol} + \Phi_{disk}\}$. We called $e$ the charge of the electron, and $\Phi_{disk}$ the flux of magnetic field outside the solenoid in the disk region comprised between $R_{sol}$ and $R_{e}$. This way the phase acquired after one electron full rotation by the electron Dirac wave function will have two terms, one depending on time and the other on the radial coordinate, $\Lambda=\Lambda(t,r)=\Lambda_{sol}(t)+\Lambda_{disk}(r)$. We could even make if deemed necessary for experiments $\Phi_{disk}$ the flux of magnetic field outside the solenoid in the disk region comprised between $R_{sol}$ and $R_{e}$ depend on the coordinate $z$ by making the magnetic field exterior to the solenoid and parallel to the solenoid axis such that $B_{disk}=B(z)$. The term corresponding to the exterior disk flux we will try to make of minimum influence in this problem because we are interested in the full inversion or time reversal \cite{FULL} and not on spacetime reflections or flips as in manuscript \cite{SCR}. We can do that by manipulating the exterior constant magnetic field that in this setup only satisfies the mission of making the electron move in circles around the solenoid symmetry axis. We rest our analysis on the idea that the electron Dirac wave equation will be gauge invariant, then necessarily the electromagnetic four-potential under whose presence the electron is moving will acquire a gauge term or stated otherwise will undergo a gauge transformation $A_{\mu} \rightarrow A_{\mu} + \Lambda_{,\mu}$. We remind ourselves that there could be different constants in the gauge transformation term, the Dirac equation minimal coupling and the exponential phase to the Dirac wavefunction according to conventions. There will be a possible non-trivial $\Lambda_{,t}$ due to the solenoid flux. We would like to have the liberty of controlling the local scalar derivative $\Lambda_{,t}$ and we can do this by changing at will the current in the solenoid coil. But most importantly we can make $\Lambda_{,t} > 0$ for pure spacetime boosts or $\Lambda_{,t} < 0$ for boosts combined with the full inversion. We can do all this increasing the current in the solenoid or decreasing the current in the solenoid, reversing the current in the solenoid, etc. $B_{disk}(t)$ would be an additional magnetic field that we include in the experimental setup in order for a charge as an electron to move in a circle of radius $R_{e}$ when the coil current $I(t)$ depends on time such that it points parallel to the axis of symmetry of the solenoid and is external to the solenoid. The need and deduction of this external magnetic field $B_{disk}(t)$ to the solenoid is explained in detail in reference \cite{KW}. The solenoid magnetic field also points in the direction parallel to the symmetry axis an has the expression $B_{sol}(t)=\mu_{o}\:n_{coil}\:I(t)$ where $n_{coil}$ is the number of coil turns per unit length and $I(t)$ is the current through the coil. See for example the problem with coil currents depending on time explained in detail in manuscript \cite{KW}. This way, the $\Lambda_{sol}(t)$ term in the expression for $\Lambda=\Lambda(t,r,z)$ will control the possibility of changing the Lorentz local scalars $C$ and $D$ in the boost or in the boost combined with the full inversion. In order for these transformations to keep the timelike or spacelike character of $V_{(1)}^{\alpha}$ and $V_{(2)}^{\alpha}$ the scalar causality condition $[(1+C)^2-D^2]>0$ must be satisfied. We repeat again that we are not dealing in this manuscript with reflections as in manuscript \cite{SCR} for which we would need $[(1+C)^2-D^2]<0$. If this condition $[(1+C)^2-D^2]>0$ is fulfilled, then we can normalize the transformed vectors $\tilde{V}_{(1)}^{\alpha}$ and $\tilde{V}_{(2)}^{\alpha}$ as follows,

\begin{eqnarray}
{\tilde{V}_{(1)}^{\alpha}
\over \sqrt{-\tilde{V}_{(1)}^{\beta}\:\tilde{V}_{(1)\beta}}}&=&
{(1+C) \over \sqrt{(1+C)^2-D^2}}
\:{V_{(1)}^{\alpha} \over \sqrt{-V_{(1)}^{\beta}\:V_{(1)\beta}}}+
{D \over \sqrt{(1+C)^2-D^2}}
\:{V_{(2)}^{\alpha} \over \sqrt{V_{(2)}^{\beta}\:V_{(2)\beta}}}\label{TN1}\\
{\tilde{V}_{(2)}^{\alpha}
\over \sqrt{\tilde{V}_{(2)}^{\beta}\:\tilde{V}_{(2)\beta}}}&=&
{D \over \sqrt{(1+C)^2-D^2}}
\:{V_{(1)}^{\alpha} \over \sqrt{-V_{(1)}^{\beta}\:V_{(1)\beta}}} +
{(1+C) \over \sqrt{(1+C)^2-D^2}}
\:{V_{(2)}^{\alpha} \over \sqrt{V_{(2)}^{\beta}\:V_{(2)\beta}}}\ .
\label{TN2}
\end{eqnarray}

The condition $[(1+C)^2-D^2]>0$ allows for two possible situations, $1+C > 0$ or $1+C < 0$, see references \cite{FULL}. For the particular case when $1+C > 0$, the transformations (\ref{TN1}-\ref{TN2}) are telling us that an electromagnetic gauge transformation on the vector field $A^{\alpha}$, that leaves invariant the electromagnetic field $f_{\mu\nu}$, generates a boost transformation on the normalized tetrad vector fields $\left({V_{(1)}^{\alpha} \over \sqrt{-V_{(1)}^{\beta}\:V_{(1)\beta}}}, {V_{(2)}^{\alpha} \over \sqrt{V_{(2)}^{\beta}\:V_{(2)\beta}}}\right)$, see reference \cite{A,IMO}. The case $1+C < 0$, represents the composition of two transformations. An inversion of the normalized tetrad vector fields $\left({V_{(1)}^{\alpha} \over \sqrt{-V_{(1)}^{\beta}\:V_{(1)\beta}}}, {V_{(2)}^{\alpha} \over \sqrt{V_{(2)}^{\beta}\:V_{(2)\beta}}}\right)$, and a boost, see equations (\ref{2TN1SC}-\ref{2TN2SC}). After n electron rotations the non-trivial phase will be $n\:\Lambda$ and the non-trivial gauge transformation of the four-electromagnetic potential of the Dirac equation will be $n\:\Lambda_{,\mu}$ and the new coefficients will be $C_{new}=n\:C_{old}$ and $D_{new}=n\:D_{old}$. See equations (\ref{COEFFC}-\ref{COEFFD}) to notice that these coefficients are linear in the gradient of $\Lambda=\Lambda(t,r,z)$. We can manage at the beginning of the experiment to create a situation where $1+C_{old} > 0$ with $C_{old} < 0$ and $\Lambda_{,t} < 0$ which is just a boost. With $C_{new}=n\:C_{old}$ if $C_{old} < 0$ there will be an n such that $1+C_{new} < 0$. If the process of the electron turning around the solenoid finally makes $1+C_{new} < 0$ then we produced by physical means a full inversion combined with a boost in such a way that we reversed future timelike oriented vectors into past timelike oriented vectors. Thus producing in a region of spacetime inside a null surface \cite{SCR} a full inversion and a time inversion as we predicted. The analysis about the existence of a null surface will be carried out in detail in a following paper \cite{NULLS}. We are not doing this analysis in this manuscript because it involves many non-trivial cases and it would deviate the main subject of research in this manuscript. It suffices to say that the null surface is defined by the equation $[(1+C)^2-D^2]=0$ and also see sections \ref{timelikenull} and \ref{sec:appIV} for a brief discussion. As a final comment we can also say that the superconducting flux quantum was actually predicted before the Aharonov and Bohm ideas by F. London in 1948 using a phenomenological model \cite{FL}.

\section{Conclusions}
\label{conclusions}

We would like to start by presenting some context for the relevance of the new group LB1. The introduction of the $SU(3)$ symmetry was relevant as to establishing an ordering principle in particle physics. We have already studied how to couple the $SU(3)$ symmetry to the gravitational field in four-dimensional curved Lorentzian spacetimes \cite{ASU3,subalgebras}. The multiplets of equal quantum numbers are translated through natural geometric elements into local multiplets of equal gravitational field. As quark physics was discovered and developed since the seventies, it was necessary to add new symmetries to the models, that resulted in the incorporation of new quantum numbers like Charm, for example. Charm is an additive quantum number like isospin $T_{3}$ and hypercharge $Y$ and the standard $T_{3}-Y$ diagrams were extended onto another third axis. Therefore, instead of the fundamental triplet we have a quartet $\{u,d,s,c\}$ as the smallest representation of the symmetry group. This last result leading to the introduction of $SU(4)$ as the new group of symmetries. We have already analyzed the coupling of the $SU(N)$ symmetry to the gravitational field \cite{ASUN}. To this end new tetrads have been introduced as we did for the $SU(3) \times SU(2) \times U(1)$ case \cite{ASU3,subalgebras}. These tetrads present outstanding properties that enable these new geometrical constructions, see for instance section \ref{electrounif} where both the gravitational and the electromagnetic field are deduced from these tetrad vectors. Many theorems have been proved regarding the isomorphic nature of these local symmetry gauge groups and tensor products of groups of local tetrad transformations. The essence of this subject is about grand field unification in four-dimensional curved Lorentzian spacetimes. The advent of the eightfold way and the quark model required the consideration of $SU(N)$ Lie groups and their representations. Because of the construction and development of Quantum Field Theories and the experimental success of its predictions including many new symmetries. Many new particles associated to $SU(N)$ representations were found and that is how these groups and their representations came into play and under consideration in fundamental physics. This process began by the beginning of the 1930s  when the isospin $SU(2)$ model was first put forward. The enormous amount of experiments involving electrodynamics, weak interactions and chromodynamics brought about the necessity to include the symmetries associated to Abelian and non-Abelian gauge groups. There is a giant literature on these subjects. We cite a limited number of books and papers that include ample sets of references \cite{DG}$^{-}$\cite{MC}. There was a parallel construction and development of the theory of General Relativity associated to gravitational phenomena in the large scale. General Relativity is another successful theory that explains a myriad of interactions involving gravitational fields like Black Holes, Gravitational Lensing, Collisions of Neutron Stars and Black Holes, Gamma Ray Bursts, and also introducing and developing theories like Quantum Gravity, etc. There is another enormous library of literature on this subject, we cite a few books and papers which include ample sets of references in relativistic astrophysics \cite{PA1}$^{-}$\cite{13} or quantum gravity \cite{JH}$^{-}$\cite{JB}. It is very well known the complexity in creating a single unified framework encompassing General Relativity and Quantum Field Theory. Tetrads have been considered in many applications during the past century \cite{ADM}$^{-}$\cite{LLIV}. We are considering in this manuscript applications based on a new kind of tetrads and group theory results associated to them. For example several experiments have been proposed in order to test the predictions emanating from these new tetrad formulations \cite{SCR,FULL,IMO,KW}. Results have been obtained in quantum mechanics, relativity, gravitation-electromagnetic unification \cite{QSMS,WSEC,EQUI}. Many results have been found in the field of fluid mechanics and relativistic astrophysics \cite{AGNS,52,symmevol,ROMP2,ENV,NSHCS,DEDM}. We focus in this review on the Abelian electromagnetic case. There is an isomorphism between the ``internal'' group of local electromagnetic gauge transformations and the local group LB1 of spacetime tetrad transformations. This ``connection'' is established through new tetrads. These new tetrads setting aside normalization factors consist of two main elements. The skeleton and the gauge vector. The gauge dependence occurs through the two local gauge vectors involved in the construction of the four tetrad vectors that allow for the study of its local gauge transformations \cite{A,ROMP,SING}. When the elements of the group LB1 are analyzed in detail we found the following, see sections \ref{sec:appII}-\ref{sec:appIV}. It is composed by the group $SO(1,1)$, the boosts, a discrete full inversion; so far all Lorentz transformations, and a new discrete transformation
given by $\Lambda^{o}_{\:\:o} = 0$, $\Lambda^{o}_{\:\:1} = 1$, $\Lambda^{1}_{\:\:o} = 1$, $\Lambda^{1}_{\:\:1} = 0$, which is clearly not a Lorentz transformation \cite{JS}$^{-}$\cite{GRSYMM} because it is a reflection plus two transformations into the intersection of the local light cone and plane one and two additional into the ``reflected'' transformation, see references \cite{A,ROMP,SING} for details and also sections \ref{timelikenull} and \ref{sec:appII}-\ref{sec:appIV}. We know from reference \cite{A} that the local group of electromagnetic gauge transformations is independently isomorphic to the local group of spatial tetrad rotations on blade two, that is $SO(2)$. This new group LB1 allows for a ``link'' between the ``internal'' transformations and ``spacetime'' transformations. This is a key issue because it has been thought for a long time  \cite{SWNG,LORNG,CMNG} that there is no relationship between the spacetime and internal groups of local transformations. The internal transformations take place in an abstract space. It has been assumed that the generators of the internal and spacetime groups commute. We proved that the locally ``internal'' group  $U(1)$ is isomorphic to a locally ``spacetime'' group of tetrad transformations on local orthogonal planes, and therefore there are assumptions made at the outset of the go-theorems \cite{SWNG,LORNG,CMNG} which are not true simply because local Lorentz transformations do not commute in general. We read from reference \cite{CMNG} ``S (the scattering matrix) is said to be Lorentz-invariant if it possesses a symmetry group locally isomorphic to the Poincar\`{e} group P.\ldots A symmetry transformation is said to be an internal symmetry transformation if it commutes with P. This implies that it acts only on particle-type indices, and has no matrix elements between particles of different four-momentum or different spin. A group composed of such transformations is called an internal symmetry group''. The local electromagnetic gauge group of transformations $U(1)$ has been proven to be isomorphic to local groups of tetrad transformations LB1 and LB2 on both the orthogonal planes one and two. The local orthogonal planes of Einstein-Maxwell stress-energy diagonalization. All vectors in these local orthogonal planes are eigenvectors of the stress-energy tensor. These local groups of transformations LB1 and LB2$=SO(2)$ are composed of Lorentz transformations and even though the LB1 special improper discrete reflection flip is not a Lorentz transformation, it is composed with this exception of spacetime Lorentz transformations, see references \cite{A,ROMP,SING}. The spacetime flip is a discrete transformation given by $\Lambda^{o}_{\:\:o} = 0$, $\Lambda^{o}_{\:\:1} = 1$, $\Lambda^{1}_{\:\:o} = 1$, $\Lambda^{1}_{\:\:1} = 0$. We notice that this discrete transformation is not a Lorentz transformation because it is a reflection. Therefore the local Lorentz group of spacetime transformations cannot commute with LB1 or LB2 since Lorentz transformations on a local plane do not commute with Lorentz transformations on another local plane necessarily. That is, the local internal groups of transformations do not necessarily commute with the local Lorentz transformations, because they are isomorphic to local groups of tetrad spacetime transformations on local orthogonal special planes. The planes of diagonalization of the stress-energy tensor, which are unique. For example in reference \cite{CMNG} we can read ``Let $G$ be a connected symmetry group of the $S$ matrix, and let the following five conditions hold: 1. (Lorentz invariance) $G$ contains a subgroup locally isomorphic to $P$ (Poincar\'{e} group). 2. (Particle-finiteness) \ldots ''. If the local gauge groups of the standard model are isomorphic to the local tetrad groups of transformations LB1 or LB2 in the Abelian electromagnetic case \cite{A,ROMP,SING,ATGU,GLAW,AEO,AMONO} or tensor products of them in the Yang-Mills general case as in references \cite{ASU3,ASUN,AYM,ATGUYM,gaugeinvmeth,A3,Sl2C,LE,Timelikenull,subalgebras} then the subgroup is $G$ itself and the no-go theorems are void of any content and incorrect. These results have a profound consequence in General Relativity, Particle Physics, Relativistic Astrophysics and Gauge Theory. To top all of these remarkable properties we have proved in this manuscript section \ref{electrounif} that these tetrads contain both the necessary information to provide the gravitational field and the necessary information to provide the electromagnetic field and the Yang-Mills field as well, in this last case see reference \cite{ATGUYM}. Analogous results were proven for the Yang-Mills cases $SU(2) \times U(1)$ see reference \cite{AYM},  $SU(3) \times SU(2) \times U(1)$ see reference \cite{ASU3}, $SL(2,C) \times SU(2) \times U(1)$  see reference \cite{Sl2C} and $SU(N) \times SU(N-1) \times \cdots \times SU(3) \times SU(2) \times U(1)$, see reference \cite{ASUN}. As an illustrative comment, in a sequel paper \cite{AYM} for the non-Abelian $SU(2)$ group, analogous isomorphism theorems were found between $SU(2)$ and the triple tensor product of LB1. We are establishing a new project for the geometrization of gauge theories, tetrad grand unification. We have already proved results for the Abelian Einstein-Maxwell case and the non-Abelian Einstein-Maxwell-Yang-Mills case in manuscripts \cite{ATGU,ATGUYM}. These new tools as we said in the abstract thus become grand unification classical objects. The perturbative classical Abelian and Yang-Mills non-Abelian cases have already been discussed in manuscripts \cite{dsmg,dsmg1,DSBYM}. In the perturbative case the local symmetries become instantaneous and evolve through time. We are simultaneously proving, and we would like to emphasize this point, that there is an isomorphism between kinematic states and gauge states of the gravitational fields locally. In paper \cite{AYM} for the non-Abelian case, we proved analogously that there is an isomorphism between kinematic states and gauge states of the gravitational fields locally. We quote from \cite{AE} ``...A second problem which at present is the subject of lively interest is the identity between the gravitational field and the electromagnetic field. The mind striving after unification of the theory cannot be satisfied that two fields should exist which, by their nature, are quite independent. A mathematically unified field theory is sought in which the gravitational field and the electromagnetic field are interpreted only as different components or manifestations of the same uniform field, the field equations where possible no longer consisting of logically mutually independent summands''.

\section{Appendix I}
\label{sec:appI}

The Levi-Civita pseudotensor can be transformed into a tensor through
the use of factors $\sqrt{-g}$, where $g$ is the determinant of the
metric tensor. We use the notation
$e_{\alpha\beta\mu\nu}=[\alpha\beta\mu\nu]$ for the
covariant components of the Levi-Civita pseudotensor in the
Minkowskian frame given in \cite{MW},

\begin{center}
$ e_{\alpha\beta\mu\nu} = \left\{ \begin{array}{ll}
				1 \:\: \mbox{if $\alpha\beta\mu\nu$ is an even permutation of 0123}\\
				-1 \:\: \mbox{if $\alpha\beta\mu\nu$ is an odd permutation of 0123}\\
				0 \:\: \mbox{if $\alpha\beta\mu\nu$ are not all different}
				    \end{array}
			    \right. $
\end{center}

It can be noticed that the signs in $e^{\alpha\beta\mu\nu}$ are going
to be opposite to the standard notation \cite{WE}.
The reason for this is that we want to keep
the compatibility with \cite{MW} where the definition
$e_{0123}=[0123]=1$ was adopted.
With these definitions we see that in a spacetime
with a metric $g_{\alpha\beta}$,

\begin{equation}
\epsilon^{\alpha\beta\mu\nu}=
{e^{\alpha\beta\mu\nu} \over  \sqrt{-g}}=
- {[\alpha\beta\mu\nu] \over \sqrt{-g}} \ ,\label{lccon}
\end{equation}

are the components of a contravariant tensor, see references \cite{LGT8} and \cite{WE,RW,LL,HS,NP,CBDW,EH,CW}.
The covariant components of (\ref{lccon}) are

\begin{equation}
\epsilon_{\alpha\beta\mu\nu}= e_{\alpha\beta\mu\nu} \sqrt{-g}=
[\alpha\beta\mu\nu] \sqrt{-g} \ ,\label{lccov}
\end{equation}

where

\begin{equation}
g_{\alpha\sigma} g_{\beta\rho}
g_{\mu\kappa} g_{\nu\lambda}\:e^{\sigma\rho\kappa\lambda}= -g\:
e_{\alpha\beta\mu\nu}\ ,
\end{equation}

is satisfied.

\section{Appendix II: The group law: The composition of two boosts}
\label{sec:appII}

According to equations (54-55) in reference \cite{A} section ``gauge geometry'', the coefficients of a local electromagnetic gauge transformation $\Lambda$ of the two vectors that span the local plane one are given by,

\begin{eqnarray}
C&=&(-Q/2)\:V_{(1)\sigma}\:\Lambda^{\sigma} / (\:V_{(2)\beta}\:
V_{(2)}^{\beta}\:)\label{COEFFCIII}\\
D&=&(-Q/2)\:V_{(2)\sigma}\:\Lambda^{\sigma} / (\:V_{(1)\beta}\:
V_{(1)}^{\beta}\:)\ .\label{COEFFDIII}
\end{eqnarray}

We would like to calculate the norm of the transformed vectors $\tilde{V}_{(1)}^{\alpha}$ and $\tilde{V}_{(2)}^{\alpha}$ that according to equations (56-57) in reference \cite{A} are given by,

\begin{eqnarray}
\tilde{V}_{(1)}^{\alpha}\:\tilde{V}_{(1)\alpha} &=&
[(1+C)^2-D^2]\:V_{(1)}^{\alpha}\:V_{(1)\alpha}\label{FPIII}\\
\tilde{V}_{(2)}^{\alpha}\:\tilde{V}_{(2)\alpha} &=&
[(1+C)^2-D^2]\:V_{(2)}^{\alpha}\:V_{(2)\alpha}\ ,\label{SPIII}
\end{eqnarray}

where the relation $V_{(1)}^{\alpha}\:V_{(1)\alpha} = -V_{(2)}^{\alpha}\:V_{(2)\alpha}$ has been used. Finally we find the transformation of the two vectors spanning the local plane one as in equations (58-59) in reference \cite{A},

\begin{eqnarray}
{\tilde{V}_{(1)}^{\alpha}
\over \sqrt{-\tilde{V}_{(1)}^{\beta}\:\tilde{V}_{(1)\beta}}}&=&
{(1+C) \over \sqrt{(1+C)^2-D^2}}
\:{V_{(1)}^{\alpha} \over \sqrt{-V_{(1)}^{\beta}\:V_{(1)\beta}}}+
{D \over \sqrt{(1+C)^2-D^2}}
\:{V_{(2)}^{\alpha} \over \sqrt{V_{(2)}^{\beta}\:V_{(2)\beta}}}\label{TN1III}\\
{\tilde{V}_{(2)}^{\alpha}
\over \sqrt{\tilde{V}_{(2)}^{\beta}\:\tilde{V}_{(2)\beta}}}&=&
{D \over \sqrt{(1+C)^2-D^2}}
\:{V_{(1)}^{\alpha} \over \sqrt{-V_{(1)}^{\beta}\:V_{(1)\beta}}} +
{(1+C) \over \sqrt{(1+C)^2-D^2}}
\:{V_{(2)}^{\alpha} \over \sqrt{V_{(2)}^{\beta}\:V_{(2)\beta}}}\ .
\label{TN2III}
\end{eqnarray}

The condition $[(1+C)^2-D^2]>0$ allows for two possible situations, $1+C > 0$ or $1+C < 0$. For the particular case when $1+C > 0$, the transformations
(\ref{TN1III}-\ref{TN2III}) are telling us that an electromagnetic gauge transformation on the vector field $A^{\alpha}$, that leaves invariant the electromagnetic field $f_{\mu\nu}$, generates a boost transformation on the normalized tetrad vector fields $\left({V_{(1)}^{\alpha} \over \sqrt{-V_{(1)}^{\beta}\:V_{(1)\beta}}}, {V_{(2)}^{\alpha} \over \sqrt{V_{(2)}^{\beta}\:V_{(2)\beta}}}\right)$. Let us consider two local consecutive electromagnetic gauge transformations $\Lambda_{1}$ and $\Lambda_{2}$ that satisfy $1+C > 0$ and the proper condition $[(1+C)^2-D^2]>0$. Let us consider the transformation of the coefficients (\ref{COEFFCIII}-\ref{COEFFDIII}) of the first local gauge electromagnetic transformation $\Lambda_{1}$ under the action of the second transformation $\Lambda_{2}$ using always the local vector tetrad transformations (\ref{TN1III}-\ref{TN2III}) and the norm of the transformed vectors (\ref{FPIII}-\ref{SPIII}),

\begin{eqnarray}
\widetilde{C}_{1} &=& (-Q/2)\:\widetilde{V}_{(1)\sigma}\:\Lambda^{\sigma}_{1} / (\:\widetilde{V}_{(2)\beta}\:
\widetilde{V}_{(2)}^{\beta}\:) = \frac{(-Q/2)}{[(1+C_{2})^2-D_{2}^2]}\: \nonumber \\ &&\times\left((1+C_{2})\:{V_{(1)\sigma}\:\Lambda^{\sigma}_{1} \over (V_{(2)}^{\beta}\:V_{(2)\beta})}+D_{2}\:{V_{(2)\sigma}\:\Lambda^{\sigma}_{1} \over (-V_{(1)}^{\beta}\:V_{(1)\beta})}\right)\label{COEFFC2III}\\
\widetilde{D}_{1} &=& (-Q/2)\:\widetilde{V}_{(2)\sigma}\:\Lambda^{\sigma}_{1} / (\:\widetilde{V}_{(1)\beta}\:
\widetilde{V}_{(1)}^{\beta}\:) = (-)\frac{(-Q/2)}{[(1+C_{2})^2-D_{2}^2]}\: \nonumber \\ &&\times\left(D_{2}
\:{V_{(1)\sigma}\:\Lambda^{\sigma}_{1} \over (V_{(2)}^{\beta}\:V_{(2)\beta})} +
(1+C_{2})\:{V_{(2)\sigma}\:\Lambda^{\sigma}_{1} \over (-V_{(1)}^{\beta}\:V_{(1)\beta})}\right)\ .\label{COEFFD2III}
\end{eqnarray}

After some calculations we finally obtain,

\begin{eqnarray}
\widetilde{C}_{1} &=&  \frac{[(1+C_{2})\:C_{1}-D_{2}\:D_{1}]}{[(1+C_{2})^2-D_{2}^2]} \label{COEFFCFIII}\\ \nonumber \\
\widetilde{D}_{1} &=&  \frac{[(1+C_{2})\:D_{1}-D_{2}\:C_{1}]}{[(1+C_{2})^2-D_{2}^2]}\ .\label{COEFFDFIII}
\end{eqnarray}

It is a matter of some algebra using equations (\ref{COEFFCFIII}-\ref{COEFFDFIII}) to find the equality,

\begin{eqnarray}
(1 + \widetilde{C}_{1})^{2} - \widetilde{D}_{1}^{2} =  \frac{[(1+C_{1}+C_{2})^{2} - (D_{1}+D_{2})^{2}]}{[(1+C_{2})^2-D_{2}^2]} \ . \label{COEFFGTIII}
\end{eqnarray}

Next we proceed to write the whole complete local sequence of electromagnetic gauge transformations $\Lambda_{1}$ and $\Lambda_{2}$ of the vectors that span the local plane one,

\begin{eqnarray}
{\widetilde{\widetilde{V}}_{(1)}^{\alpha}
\over \sqrt{-\widetilde{\widetilde{V}}_{(1)}^{\beta}\:\widetilde{\widetilde{V}}_{(1)\beta}}}&=&
{(1+\widetilde{C}_{1}) \over \sqrt{(1+\widetilde{C}_{1})^2-\widetilde{D}_{1}^2}}
\:{\widetilde{V}_{(1)}^{\alpha} \over \sqrt{-\widetilde{V}_{(1)}^{\beta}\:\widetilde{V}_{(1)\beta}}}+
{\widetilde{D}_{1} \over \sqrt{(1+\widetilde{C}_{1})^2-\widetilde{D}_{1}^2}}
\:{\widetilde{V}_{(2)}^{\alpha} \over \sqrt{\widetilde{V}_{(2)}^{\beta}\:\widetilde{V}_{(2)\beta}}}\label{TN1SIII}\\
{\widetilde{\widetilde{V}}_{(2)}^{\alpha}
\over \sqrt{\widetilde{\widetilde{V}}_{(2)}^{\beta}\:\widetilde{\widetilde{V}}_{(2)\beta}}}&=&
{\widetilde{D}_{1} \over \sqrt{(1+\widetilde{C}_{1})^2-\widetilde{D}_{1}^2}}
\:{\widetilde{V}_{(1)}^{\alpha} \over \sqrt{-\widetilde{V}_{(1)}^{\beta}\:\widetilde{V}_{(1)\beta}}} +
{(1+\widetilde{C}_{1}) \over \sqrt{(1+\widetilde{C}_{1})^2-\widetilde{D}_{1}^2}}
\:{\widetilde{V}_{(2)}^{\alpha} \over \sqrt{\widetilde{V}_{(2)}^{\beta}\:\widetilde{V}_{(2)\beta}}} \ ,
\label{TN2SIII}
\end{eqnarray}

where,

\begin{eqnarray}
{\tilde{V}_{(1)}^{\alpha}
\over \sqrt{-\tilde{V}_{(1)}^{\beta}\:\tilde{V}_{(1)\beta}}}&=&
{(1+C_{2}) \over \sqrt{(1+C_{2})^2-D_{2}^2}}
\:{V_{(1)}^{\alpha} \over \sqrt{-V_{(1)}^{\beta}\:V_{(1)\beta}}}+
{D_{2} \over \sqrt{(1+C_{2})^2-D_{2}^2}}
\:{V_{(2)}^{\alpha} \over \sqrt{V_{(2)}^{\beta}\:V_{(2)\beta}}}\label{TN1TIII}\\
{\tilde{V}_{(2)}^{\alpha}
\over \sqrt{\tilde{V}_{(2)}^{\beta}\:\tilde{V}_{(2)\beta}}}&=&
{D_{2} \over \sqrt{(1+C_{2})^2-D_{2}^2}}
\:{V_{(1)}^{\alpha} \over \sqrt{-V_{(1)}^{\beta}\:V_{(1)\beta}}} +
{(1+C_{2}) \over \sqrt{(1+C_{2})^2-D_{2}^2}}
\:{V_{(2)}^{\alpha} \over \sqrt{V_{(2)}^{\beta}\:V_{(2)\beta}}}\ .
\label{TN2TIII}
\end{eqnarray}

After we make use of equations (\ref{COEFFCFIII}-\ref{COEFFGTIII}) we can write after some algebraic calculations the final expressions for the sequence of two local electromagnetic gauge transformations $\Lambda_{1}$ and $\Lambda_{2}$ of the two vectors that span the local plane one,

\begin{eqnarray}
{\widetilde{\widetilde{V}}_{(1)}^{\alpha}
\over \sqrt{-\widetilde{\widetilde{V}}_{(1)}^{\beta}\:\widetilde{\widetilde{V}}_{(1)\beta}}}&=&
{[1+C_{1}+C_{2}] \over \sqrt{(1+C_{1}+C_{2})^2-(D_{1}+D_{2})^2}}
\:{V_{(1)}^{\alpha} \over \sqrt{-V_{(1)}^{\beta}\:V_{(1)\beta}}}+ \nonumber \\
&&{[D_{1}+D_{2}] \over \sqrt{(1+C_{1}+C_{2})^2-(D_{1}+D_{2})^2}}
\:{V_{(2)}^{\alpha} \over \sqrt{V_{(2)}^{\beta}\:V_{(2)\beta}}} \label{TN1FINALIII}\\
{\widetilde{\widetilde{V}}_{(2)}^{\alpha}
\over \sqrt{\widetilde{\widetilde{V}}_{(2)}^{\beta}\:\widetilde{\widetilde{V}}_{(2)\beta}}}&=&
{[D_{1}+D_{2}] \over \sqrt{(1+C_{1}+C_{2})^2-(D_{1}+D_{2})^2}}
\:{V_{(1)}^{\alpha} \over \sqrt{-V_{(1)}^{\beta}\:V_{(1)\beta}}} + \nonumber \\
&&{[1+C_{1}+C_{2}] \over \sqrt{(1+C_{1}+C_{2})^2-(D_{1}+D_{2})^2}}
\:{V_{(2)}^{\alpha} \over \sqrt{V_{(2)}^{\beta}\:V_{(2)\beta}}} \ .
\label{TN2FINALIII}
\end{eqnarray}

As we can see the local transformation of the two vectors that span the local plane one by two consecutive local electromagnetic gauge transformations that satisfy $1+C > 0$ and the proper condition $[(1+C)^2-D^2]>0$, is found to be equal to the addition of both as it was expected from an Abelian group. Therefore in this first case of two boosts composed and generated by $\Lambda_{1}$ and $\Lambda_{2}$ if we call $L_{boost}(C,D)$ the local Lorentz transformation generated by one boost, then for the composition of two boosts we found through equations (\ref{TN1SIII}-\ref{TN2FINALIII}) the relation $L_{boost}(C_{1},D_{1})\:\:\:\:L_{boost}(C_{2},D_{2}) = L_{boost}(C_{1}+C_{2},D_{1}+D_{2})$ as expected. The coefficients $(\widetilde{C}_{1},\widetilde{D}_{1})$ are given in equations (\ref{COEFFCFIII}-\ref{COEFFDFIII}) and they are auxiliary quantities in the intermediate calculation $L_{boost}(\widetilde{C}_{1},\widetilde{D}_{1})\:\:\:\:L_{boost}(C_{2},D_{2}) = L_{boost}(C_{1}+C_{2},D_{1}+D_{2})$. It is worth mentioning that the composition of a transformation of the kind (58-59) as given in reference \cite{A} with $[(1+C)^2-D^2]>0$ and $1+C > 0$ and a transformation of the kind (60-61) as given in reference \cite{A} with $[(1+C)^2-D^2]>0$ and $1+C < 0$ would proceed similarly to the composition of two boosts of the kind (58-59) as given in reference \cite{A}. In fact the composition of two transformations of the kind (60-61) as given in reference \cite{A} with $[(1+C)^2-D^2]>0$ and $1+C < 0$ would proceed similarly to the composition of two boosts.  However we have to be very careful about the final result because sometimes the result might be a special improper transformation as exemplified in the next set of examples below. Nonetheless, the group law will always hold, see reference \cite{GLAW}.

\subsubsection{Example}
\label{example-pp}

Let us consider a first proper local gauge transformation with parameters $1+C_{1}=2/\sqrt{3}$ and $D_{1}=1/\sqrt{3}$, a boost. Let us consider a second local proper gauge transformation with parameters $1+C_{2}=5$ and $D_{2}=2\:\sqrt{6}$, another boost. The group law (\ref{TN1FINALIII}-\ref{TN2FINALIII}) will determine that the final composed transformation is given by the parameters $1+C_{1}+C_{2}=5+2/\sqrt{3}$ and $D_{1}+D_{2}=2\:\sqrt{6}+\frac{1}{\sqrt{3}}$. The composition of two proper transformations gives an proper boost transformation, since $(1+C_{1}+C_{2})^{2} - (D_{1}+D_{2})^{2} > 0$. All belong to LB1.

We have to make an observation at this point. Let us consider a first proper local gauge transformation with parameters $1+C_{1}=2/\sqrt{3}$ and $D_{1}=1/\sqrt{3}$, a boost. Let us consider a second local proper gauge transformation with parameters $1+C_{2}=-2/\sqrt{3}$ and $D_{2}=1/\sqrt{3}$, a boost composed with a full inversion since $1+C_{2}<0$, which is nonetheless a proper transformation with determinant equal to one. The group law (\ref{TN1FINALIII}-\ref{TN2FINALIII}) will determine that the final composed transformation is given by the parameters $1+C_{1}+C_{2}=-1$ and $D_{1}+D_{2}=2/\sqrt{3}$. The composition of two proper transformations gives an special improper  transformation, since $(1+C_{1}+C_{2})^{2} - (D_{1}+D_{2})^{2} < 0$. The reason is that one of the proper transformations is composed with a full inversion in this particular example. However, the group law is satisfied because the proof is identical to the above in this section \ref{sec:appII}. The reason is that in equation (\ref{COEFFGTIII}) both sides would be now negative in this particular example and when expressing equations (\ref{TN1SIII}-\ref{TN2SIII}) in the denominators of all terms there would be a $\sqrt{-(1+\widetilde{C}_{1})^2+\widetilde{D}_{1}^2}$ instead of $\sqrt{(1+\widetilde{C}_{1})^2-\widetilde{D}_{1}^2}$. In the denominator of all terms in equations (\ref{TN1FINALIII}-\ref{TN2FINALIII}) there would be a $\sqrt{-(1+C_{1}+C_{2})^2+(D_{1}+D_{2})^2}$ instead of $\sqrt{(1+C_{1}+C_{2})^2-(D_{1}+D_{2})^2}$. This is a particular example we remind ourselves and the rest of the analysis would be analogous. Not in all compositions of boosts and boosts + full inversion the same situation will arise necessarily. In the end all of the transformations involved belong in LB1.

\subsection{Direct test of the group law}
\label{dtestgl}

We will study the composition of one boost with another second transformation a boost composed with a full inversion. Please remember that the example in section \ref{example-pp} and the example in this section are not the same. We will study how two consecutive transformations such as a direct and the inverse will result in the identity and not minus the identity as we might suspect in the following reasoning and deduction.
Let us consider the equations that start at (\ref{COEFFCIII}-\ref{COEFFDIII}) and end at equations (\ref{TN1FINALIII}-\ref{TN2FINALIII}). Let us consider next two local consecutive electromagnetic gauge transformations $\Lambda_{1}$ and $\Lambda_{2}$ that satisfy $1+C_{1} > 0$, $1+C_{2} < 0$ and the proper condition $[(1+C)^2-D^2]>0$ for both. Let us consider the transformation of the coefficients of the first local gauge electromagnetic transformation $\Lambda_{1}$ under the action of the second transformation $\Lambda_{2}$ using always the local vector tetrad transformations (\ref{TN1III}-\ref{TN2III}) and the norm of the transformed vectors (\ref{FPIII}-\ref{SPIII}),

\begin{eqnarray}
\widetilde{C}_{1} &=& (-Q/2)\:\widetilde{V}_{(1)\sigma}\:\Lambda^{\sigma}_{1} / (\:\widetilde{V}_{(2)\beta}\:
\widetilde{V}_{(2)}^{\beta}\:) = \frac{(-Q/2)}{[(1+C_{2})^2-D_{2}^2]}\: \nonumber \\ &&\times\left((1+C_{2})\:{V_{(1)\sigma}\:\Lambda^{\sigma}_{1} \over (V_{(2)}^{\beta}\:V_{(2)\beta})}+D_{2}\:{V_{(2)\sigma}\:\Lambda^{\sigma}_{1} \over (-V_{(1)}^{\beta}\:V_{(1)\beta})}\right)\label{COEFFC2}\\
\widetilde{D}_{1} &=& (-Q/2)\:\widetilde{V}_{(2)\sigma}\:\Lambda^{\sigma}_{1} / (\:\widetilde{V}_{(1)\beta}\:
\widetilde{V}_{(1)}^{\beta}\:) = (-)\frac{(-Q/2)}{[(1+C_{2})^2-D_{2}^2]}\: \nonumber \\ &&\times\left(D_{2}
\:{V_{(1)\sigma}\:\Lambda^{\sigma}_{1} \over (V_{(2)}^{\beta}\:V_{(2)\beta})} +
(1+C_{2})\:{V_{(2)\sigma}\:\Lambda^{\sigma}_{1} \over (-V_{(1)}^{\beta}\:V_{(1)\beta})}\right)\ .\label{COEFFD2}
\end{eqnarray}

After some calculations we finally obtain,

\begin{eqnarray}
\widetilde{C}_{1} &=&  \frac{[(1+C_{2})\:C_{1}-D_{2}\:D_{1}]}{[(1+C_{2})^2-D_{2}^2]} \label{COEFFCF}\\ \nonumber \\
\widetilde{D}_{1} &=&  \frac{[(1+C_{2})\:D_{1}-D_{2}\:C_{1}]}{[(1+C_{2})^2-D_{2}^2]}\ .\label{COEFFDF}
\end{eqnarray}

It is a matter of some algebra using equations (\ref{COEFFCF}-\ref{COEFFDF}) to find the equality,

\begin{eqnarray}
(1 + \widetilde{C}_{1})^{2} - \widetilde{D}_{1}^{2} =  \frac{[(1+C_{1}+C_{2})^{2} - (D_{1}+D_{2})^{2}]}{[(1+C_{2})^2-D_{2}^2]} \ . \label{COEFFGT}
\end{eqnarray}

We will assume for our particular situation that both sides of equation (\ref{COEFFGT}) are positive. We will show an example with these properties. Next we proceed to write the whole complete local sequence of electromagnetic gauge transformations $\Lambda_{1}$ and $\Lambda_{2}$ of the vectors that span the local plane one,

\begin{eqnarray}
{\widetilde{\widetilde{V}}_{(1)}^{\alpha}
\over \sqrt{-\widetilde{\widetilde{V}}_{(1)}^{\beta}\:\widetilde{\widetilde{V}}_{(1)\beta}}}&=&
{(1+\widetilde{C}_{1}) \over \sqrt{(1+\widetilde{C}_{1})^2-\widetilde{D}_{1}^2}}
\:{\widetilde{V}_{(1)}^{\alpha} \over \sqrt{-\widetilde{V}_{(1)}^{\beta}\:\widetilde{V}_{(1)\beta}}}+
{\widetilde{D}_{1} \over \sqrt{(1+\widetilde{C}_{1})^2-\widetilde{D}_{1}^2}}
\:{\widetilde{V}_{(2)}^{\alpha} \over \sqrt{\widetilde{V}_{(2)}^{\beta}\:\widetilde{V}_{(2)\beta}}}\label{TN1S}\\
{\widetilde{\widetilde{V}}_{(2)}^{\alpha}
\over \sqrt{\widetilde{\widetilde{V}}_{(2)}^{\beta}\:\widetilde{\widetilde{V}}_{(2)\beta}}}&=&
{\widetilde{D}_{1} \over \sqrt{(1+\widetilde{C}_{1})^2-\widetilde{D}_{1}^2}}
\:{\widetilde{V}_{(1)}^{\alpha} \over \sqrt{-\widetilde{V}_{(1)}^{\beta}\:\widetilde{V}_{(1)\beta}}} +
{(1+\widetilde{C}_{1}) \over \sqrt{(1+\widetilde{C}_{1})^2-\widetilde{D}_{1}^2}}
\:{\widetilde{V}_{(2)}^{\alpha} \over \sqrt{\widetilde{V}_{(2)}^{\beta}\:\widetilde{V}_{(2)\beta}}} \ ,
\label{TN2S}
\end{eqnarray}

where,

\begin{eqnarray}
{\tilde{V}_{(1)}^{\alpha}
\over \sqrt{-\tilde{V}_{(1)}^{\beta}\:\tilde{V}_{(1)\beta}}}&=&
{(1+C_{2}) \over \sqrt{(1+C_{2})^2-D_{2}^2}}
\:{V_{(1)}^{\alpha} \over \sqrt{-V_{(1)}^{\beta}\:V_{(1)\beta}}}+
{D_{2} \over \sqrt{(1+C_{2})^2-D_{2}^2}}
\:{V_{(2)}^{\alpha} \over \sqrt{V_{(2)}^{\beta}\:V_{(2)\beta}}}\label{TN1T}\\
{\tilde{V}_{(2)}^{\alpha}
\over \sqrt{\tilde{V}_{(2)}^{\beta}\:\tilde{V}_{(2)\beta}}}&=&
{D_{2} \over \sqrt{(1+C_{2})^2-D_{2}^2}}
\:{V_{(1)}^{\alpha} \over \sqrt{-V_{(1)}^{\beta}\:V_{(1)\beta}}} +
{(1+C_{2}) \over \sqrt{(1+C_{2})^2-D_{2}^2}}
\:{V_{(2)}^{\alpha} \over \sqrt{V_{(2)}^{\beta}\:V_{(2)\beta}}}\ .
\label{TN2T}
\end{eqnarray}

After we make use of equations (\ref{COEFFCF}-\ref{COEFFGT}) we can write after some algebraic calculations the final expressions for the sequence of two local electromagnetic gauge transformations $\Lambda_{1}$ and $\Lambda_{2}$ of the two vectors that span the local plane one,

\begin{eqnarray}
{\widetilde{\widetilde{V}}_{(1)}^{\alpha}
\over \sqrt{-\widetilde{\widetilde{V}}_{(1)}^{\beta}\:\widetilde{\widetilde{V}}_{(1)\beta}}}&=&
{[1+C_{1}+C_{2}] \over \sqrt{(1+C_{1}+C_{2})^2-(D_{1}+D_{2})^2}}
\:{V_{(1)}^{\alpha} \over \sqrt{-V_{(1)}^{\beta}\:V_{(1)\beta}}}+ \nonumber \\
&&{[D_{1}+D_{2}] \over \sqrt{(1+C_{1}+C_{2})^2-(D_{1}+D_{2})^2}}
\:{V_{(2)}^{\alpha} \over \sqrt{V_{(2)}^{\beta}\:V_{(2)\beta}}} \label{TN1FINAL}\\
{\widetilde{\widetilde{V}}_{(2)}^{\alpha}
\over \sqrt{\widetilde{\widetilde{V}}_{(2)}^{\beta}\:\widetilde{\widetilde{V}}_{(2)\beta}}}&=&
{[D_{1}+D_{2}] \over \sqrt{(1+C_{1}+C_{2})^2-(D_{1}+D_{2})^2}}
\:{V_{(1)}^{\alpha} \over \sqrt{-V_{(1)}^{\beta}\:V_{(1)\beta}}} + \nonumber \\
&&{[1+C_{1}+C_{2}] \over \sqrt{(1+C_{1}+C_{2})^2-(D_{1}+D_{2})^2}}
\:{V_{(2)}^{\alpha} \over \sqrt{V_{(2)}^{\beta}\:V_{(2)\beta}}} \ .
\label{TN2FINAL}
\end{eqnarray}

As we can see the local transformation of the two vectors that span the local plane one by two consecutive local electromagnetic gauge transformations that satisfy the proper condition $[(1+C)^2-D^2]>0$, is found to be equal to the addition of both as it was expected from an Abelian group even though highly not-trivial because of the relations (\ref{COEFFCF}-\ref{COEFFDF}).

Let us next consider for our particular example that $C_{1}=C>0$, $D_{1}=0$, $C_{2}=C^{inv}=-C_{1}=-C<0$ and $D_{2}=0$. We also know from our assumptions in the beginning of section XIII in reference \cite{SING} that for $n$ large enough $1+C_{1}=1+C>0$ a boost and $1+C_{2}=1-C<0$ a boost composed with a full inversion. That is, if we consider instead of the local gauge transformation $\Lambda$ the local gauge transformation $n\:\Lambda$ for $n$ a natural number, the coefficients of the transformation will be given for example by $C_{1,n>1}=n\:C_{1,n=1}$ and $D_{1,n>1}=n\:D_{1,n=1}$ and similar for $C_{2,n>1}=n\:C_{2,n=1}$ and $D_{2,n>1}=n\:D_{2,n=1}$ . If we use equations (\ref{COEFFCF}-\ref{COEFFDF}) we find,

\begin{eqnarray}
\widetilde{C}_{1} &=&  \frac{[(1+C_{2})\:C_{1}-D_{2}\:D_{1}]}{[(1+C_{2})^2-D_{2}^2]} = \frac{C_{1}}{1-C_{1}} \label{COEFFCFRNE}\\ \nonumber \\
\widetilde{D}_{1} &=&  \frac{[(1+C_{2})\:D_{1}-D_{2}\:C_{1}]}{[(1+C_{2})^2-D_{2}^2]} = 0 \ .\label{COEFFDFRNE}
\end{eqnarray}

We also find

\begin{eqnarray}
(1 + \widetilde{C}_{1})^{2} - \widetilde{D}_{1}^{2} =  \frac{[(1+C_{1}+C_{2})^{2} - (D_{1}+D_{2})^{2}]}{[(1+C_{2})^2-D_{2}^2]} = \frac{1}{(1-C_{1})^{2}}\ . \label{COEFFGTRNE}
\end{eqnarray}

When we use in sequence equations (\ref{TN1S}-\ref{TN2S}) and (\ref{TN1T}-\ref{TN2T}) we find,

\begin{eqnarray}
{\widetilde{\widetilde{V}}_{(1)}^{\alpha}
\over \sqrt{-\widetilde{\widetilde{V}}_{(1)}^{\beta}\:\widetilde{\widetilde{V}}_{(1)\beta}}}&=&
{|1-C_{1}| \over (1-C_{1})}
\:{\widetilde{V}_{(1)}^{\alpha} \over \sqrt{-\widetilde{V}_{(1)}^{\beta}\:\widetilde{V}_{(1)\beta}}} \label{TN1SRNE}\\
{\widetilde{\widetilde{V}}_{(2)}^{\alpha}
\over \sqrt{\widetilde{\widetilde{V}}_{(2)}^{\beta}\:\widetilde{\widetilde{V}}_{(2)\beta}}}&=&
{|1-C_{1}| \over (1-C_{1})}
\:{\widetilde{V}_{(2)}^{\alpha} \over \sqrt{\widetilde{V}_{(2)}^{\beta}\:\widetilde{V}_{(2)\beta}}} \ ,
\label{TN2SRNE}
\end{eqnarray}

where,

\begin{eqnarray}
{\tilde{V}_{(1)}^{\alpha}
\over \sqrt{-\tilde{V}_{(1)}^{\beta}\:\tilde{V}_{(1)\beta}}}&=&
{(1-C_{1}) \over |1-C_{1}|}
\:{V_{(1)}^{\alpha} \over \sqrt{-V_{(1)}^{\beta}\:V_{(1)\beta}}} \label{TN1TRNE}\\
{\tilde{V}_{(2)}^{\alpha}
\over \sqrt{\tilde{V}_{(2)}^{\beta}\:\tilde{V}_{(2)\beta}}}&=&
{(1-C_{1}) \over |1-C_{1}|}
\:{V_{(2)}^{\alpha} \over \sqrt{V_{(2)}^{\beta}\:V_{(2)\beta}}}\ .
\label{TN2TRNE}
\end{eqnarray}

We have proved in the end that the apparent contradiction of having one transformation with $1+C>0$ to be the identity while the inverse with $1+C^{inv}=1-C<0$ is minus the identity as far as it regards the group law is not a true inconsistence because the composition transformation is not just the product of these two matrices giving minus the identity but a far more involved and non-trivial composition as presented through the whole work in this section. From equations (\ref{TN1SRNE}-\ref{TN2SRNE}) and (\ref{TN1TRNE}-\ref{TN2TRNE}) we have proved that the composition is consistent with the direct result obtained from equations (\ref{TN1FINAL}-\ref{TN2FINAL}) where $1+C_{1}+C_{2}=1$ and $D_{1}+D_{2}=0$. That is, the identity in both separate calculations, see references \cite{SING,GLAW}.

\subsection{Reverse test of the group law}
\label{dtestgl}

Let us next consider the reverse order in gauge tetrad transformations for our particular example such that $C_{1}=-C<0$, $D_{1}=0$, $C_{2}=-C^{inv}=-C_{1}=C>0$ and $D_{2}=0$. We also know from our assumptions in the beginning of section XIII in manuscript \cite{SING} that for $n$ large enough natural number $1+C_{1}=1-C<0$ a boost composed with a full inversion and $1+C_{2}=1+C>0$ a boost as reasoned just above. If we use equations (\ref{COEFFCF}-\ref{COEFFDF}) we find,

\begin{eqnarray}
\widetilde{C}_{1} &=&  \frac{[(1+C_{2})\:C_{1}-D_{2}\:D_{1}]}{[(1+C_{2})^2-D_{2}^2]} = \frac{C_{1}}{1-C_{1}} \label{COEFFCFRNERO}\\ \nonumber \\
\widetilde{D}_{1} &=&  \frac{[(1+C_{2})\:D_{1}-D_{2}\:C_{1}]}{[(1+C_{2})^2-D_{2}^2]} = 0 \ .\label{COEFFDFRNERO}
\end{eqnarray}

We also find

\begin{eqnarray}
(1 + \widetilde{C}_{1})^{2} - \widetilde{D}_{1}^{2} =  \frac{[(1+C_{1}+C_{2})^{2} - (D_{1}+D_{2})^{2}]}{[(1+C_{2})^2-D_{2}^2]} = \frac{1}{(1-C_{1})^{2}}\ . \label{COEFFGTRNERO}
\end{eqnarray}

When we use in sequence equations (\ref{TN1S}-\ref{TN2S}) and (\ref{TN1T}-\ref{TN2T}) we find,

\begin{eqnarray}
{\widetilde{\widetilde{V}}_{(1)}^{\alpha}
\over \sqrt{-\widetilde{\widetilde{V}}_{(1)}^{\beta}\:\widetilde{\widetilde{V}}_{(1)\beta}}}&=&
{|1-C_{1}| \over (1-C_{1})}
\:{\widetilde{V}_{(1)}^{\alpha} \over \sqrt{-\widetilde{V}_{(1)}^{\beta}\:\widetilde{V}_{(1)\beta}}} \label{TN1SRNERO}\\
{\widetilde{\widetilde{V}}_{(2)}^{\alpha}
\over \sqrt{\widetilde{\widetilde{V}}_{(2)}^{\beta}\:\widetilde{\widetilde{V}}_{(2)\beta}}}&=&
{|1-C_{1}| \over (1-C_{1})}
\:{\widetilde{V}_{(2)}^{\alpha} \over \sqrt{\widetilde{V}_{(2)}^{\beta}\:\widetilde{V}_{(2)\beta}}} \ ,
\label{TN2SRNERO}
\end{eqnarray}

where,

\begin{eqnarray}
{\tilde{V}_{(1)}^{\alpha}
\over \sqrt{-\tilde{V}_{(1)}^{\beta}\:\tilde{V}_{(1)\beta}}}&=&
{(1-C_{1}) \over |1-C_{1}|}
\:{V_{(1)}^{\alpha} \over \sqrt{-V_{(1)}^{\beta}\:V_{(1)\beta}}} \label{TN1TRNERO}\\
{\tilde{V}_{(2)}^{\alpha}
\over \sqrt{\tilde{V}_{(2)}^{\beta}\:\tilde{V}_{(2)\beta}}}&=&
{(1-C_{1}) \over |1-C_{1}|}
\:{V_{(2)}^{\alpha} \over \sqrt{V_{(2)}^{\beta}\:V_{(2)\beta}}}\ .
\label{TN2TRNERO}
\end{eqnarray}

We notice that the difference between equations (\ref{TN1SRNE}-\ref{TN2SRNE}) plus (\ref{TN1TRNE}-\ref{TN2TRNE}) on one hand and equations (\ref{TN1SRNERO}-\ref{TN2SRNERO}) plus (\ref{TN1TRNERO}-\ref{TN2TRNERO}) on the other hand is that in the first case ${|1-C_{1}| \over (1-C_{1})}=-1$ while in the second case ${|1-C_{1}| \over (1-C_{1})}=+1$. We have proved in the end that the apparent contradiction of having one transformation with $1-C<0$ to be minus the identity while the inverse with $1-C^{inv}=1+C>0$ to be the identity as far as it regards the group law is not a true inconsistence because the composition transformation is not just the product of these two matrices giving minus the identity but a far more involved and non-trivial composition as presented through the whole work in this section. From equations (\ref{TN1SRNERO}-\ref{TN2SRNERO}) and (\ref{TN1TRNERO}-\ref{TN2TRNERO}) we have proved that the composition is consistent with the direct result obtained from equations (\ref{TN1FINAL}-\ref{TN2FINAL}) where $1+C_{1}+C_{2}=1$ and $D_{1}+D_{2}=0$. That is, the identity in both separate calculations also for the group law with the transformations in reverse order.

\section{Appendix III: Kernel of the mapping}
\label{sec:appIII}

The proof to the following theorems can be found in references \cite{A,ROMP}. In order to summarize all the results in this section which we will need in the subsequent sections we state the following,

\begin{itemize}

\item In reference \cite{ROMP} it was found that the Kernel of the map between the local group of electromagnetic gauge transformations and the local group of tetrad transformations in the proper sector in the local blade one is just composed by the group $PGB2$, where $PGB2=\{\Lambda / \Lambda^{\mu} \in \mbox{local Plane 2} \}$ is the set of pure gauge in blade two, as long as the choice for tetrad gauge vector is not pure gauge $X^{\mu} \neq \Lambda^{\mu}$ or the pure gauge multiplied by a local scalar $X^{\mu} \neq \frac{1}{C}\:\Lambda^{\mu}$ with $1+C>0$ which is equivalent to pure gauge for the skeleton-gauge vector tetrad structure on plane one. The group $PGB2$ is of measure zero and it is just gradients of scalars in a local plane in a four-dimensional spacetime. There is an isomorphism between the group of local electromagnetic gauge transformations minus the set $PGB2$ and LB1. The notation for the local scalar $1+C$ is the same as in section \ref{localbladeone}. Please see reference \cite{ROMP} for the whole analysis.

\item In reference \cite{ROMP} it was found that the Kernel of the map between the local group of electromagnetic gauge transformations and the local group of tetrad transformations on the local blade two is just composed by the group $PGB1$, where $PGB1=\{\Lambda / \Lambda^{\mu} \in \mbox{local Plane 1} \}$ is the set of pure gauge in blade one, as long as the choice for tetrad gauge vector is not pure gauge $Y^{\mu} \neq \ast \Lambda^{\mu}$ or the pure gauge multiplied by a local scalar $X^{\mu} \neq \frac{1}{N}\:\ast \Lambda^{\mu}$ with $1+N>0$ which is equivalent to pure gauge for the skeleton-gauge vector tetrad structure on plane two. The group $PGB1$ is of measure zero and it is just gradients of scalars in a local plane in a four-dimensional spacetime. There is an isomorphism between the group of local electromagnetic gauge transformations minus the set $PGB1$ and LB2. The notation for the local scalar $1+N$ is the same as in section \ref{localbladeone}. Please see reference \cite{ROMP} for the whole analysis.

\item In reference \cite{ROMP} it was found that the map between $U(1)$ and $LB1 \otimes LB2$ is an isomorphism. The Kernel of this map will be just constant gauge transformations.

\end{itemize}

The mapping between $U(1)$ and $LB1 \otimes LB2$ will be an isomorphism. In the general sense this isomorphism is piecewise. Because we have in the local plane one boosts which are hyperbolic rotations, boosts composed with full inversions which are the composition of two reflections, boosts composed with spacetime reflections and boosts composed with full inversions and spacetime reflections. In the local plane two we have spatial rotations. It is in a general sense a piecewise isomorphism, see reference \cite{ROMP}.

%In the next section we will study the more subtle mappings given by equations (60-61), (64-65) and (66-67) in reference \cite{A}.

\subsubsection{PGB1 isomorphic to PGB2}
\label{pgiso}

We would like to know if we can establish a one to one relationship between the groups $PGB1$ and $PGB2$. The answer is affirmative and please see the proof in manuscript \cite{ROMP}.

\section{Appendix IV: Infinity limits}
\label{sec:appIV}

Let us consider the plane with rectangular coordinates $(X,Y)$ and on this plane we will also consider the hyperbola $X^{2}-Y^{2}=1$. But for simplicity we will only consider the upper half branch $X\geq0$ and $Y\geq0$. We intend to study the possible existence of a map between this upper half branch and the quadrant $X\geq0$ and $Y\geq0$ for the circle $X^{2}+Y^{2}=1$. The method that we will use involves the unit sphere $S^{2}$ and the stereographic projection through the north pole. We will study the hyperbola and the circle on a plane that cuts the sphere through the equator in such a way that the circle is the equator of the unit sphere $S^{2}$. Let us introduce the coordinates of the stereographic projections for the unit sphere $S^{2}$. The local 2-sphere is defined through $\sum_{i=1}^{3} x_{i}^{2} = 1$ where the $x_{i},\: i=1 \cdots 3$ are local coordinates. Following closely chapter III in \cite{CBDW} an in order to construct an atlas we let $P$ and $Q$ be the north and south poles respectively. Let $U=S^{2}-{P}$ and $V=S^{2}-{Q}$, let g and h be the stereographic projections of the poles $P$ and $Q$ on the plane $x_{3}=0$,

\begin{center}
$\: g: U \rightarrow \Re^{2} \:\:\: by  \:\:\:  y_{i} = x_{i} / (1-x_{3}) \:\:\: for  \:\:\:  i=1 \cdots 2$
\end{center}

\begin{center}
$\: h: V \rightarrow \Re^{2} \:\:\: by  \:\:\: z_{i} = x_{i} / (1+x_{3}) \:\:\: for  \:\:\: i=1 \cdots 2$
\end{center}

See reference \cite{CBDW} for the proof that this is an atlas. Let us parameterize the upper half branch of the hyperbola under consideration by $(\cosh(t),\sinh(t))$ with $t:0 \rightarrow +\infty$. In turn we will parameterize the unit circle on the same plane quadrant with $(\cos(\varphi),\sin(\varphi))$ and with $\varphi:0 \rightarrow \pi/2$. Then, if the coordinates on the sphere are $(x,y,z)$ and the coordinates on the plane are $(X,Y)$ the relationship between them through the stereographic projection of the north pole will be,

\begin{center}
$(X,Y) = ({x \over 1-z},{ y \over 1-z})$
\end{center}

\begin{center}
$(x,y,z) = ({2\:X \over 1+X^{2}+Y^{2}},{ 2\:Y \over 1+X^{2}+Y^{2}}, {-1+X^{2}+Y^{2} \over 1+X^{2}+Y^{2}})$
\end{center}

If in the second of these two stereographic coordinate relationships we set $(X=\cosh(t),Y=\sinh(t))$ we obtain,

\begin{center}
$(x,y,z) = ({2\:\cosh(t) \over 1+\cosh^{2}(t)+\sinh^{2}(t)},{ 2\:\sinh(t) \over 1+\cosh^{2}(t)+\sinh^{2}(t)}, {-1+\cosh^{2}(t)+\sinh^{2}(t) \over 1+\cosh^{2}(t)+\sinh^{2}(t)})$
\end{center}

We might also notice that $\cosh^{2}(t)+\sinh^{2}(t)=\cosh(2t)$. Finally we establish the following mapping,

\begin{eqnarray}
\cos(\varphi) &=& {2\:\cosh(t) \over 1+\cosh^{2}(t)+\sinh^{2}(t)} = {2\:\cosh(t) \over 1+\cosh(2t)} \label{stereomap}
\end{eqnarray}

Within the range $t:0 \rightarrow +\infty$ and $\varphi:0 \rightarrow (\pi/2)^{-}$ this map is truly and isomorphism. We can immediately notice that the infinity value for the coordinate t will correspond to the north pole and the isomorphic map will not reach $\pi/2$ even though $\pi/2$ is an accumulation point. Then the issue of the point at infinity arises. It is evident that it must be added in order to reach the point $\cos\varphi=\pi/2$. This map is telling us that the upper branch for $X\geq0$ and $Y\geq0$ maps into the first quadrant of the circle except for the point $\varphi=\pi/2$ which is an accumulation point. It happens that the point at infinity is the point in the asymptote corresponding to the future light cone found in sections \ref{diffeq}-\ref{diffeqrn}-\ref{geneq}. There is a unique inhomogeneous solution as in sections \ref{diffeq}-\ref{diffeqrn}-\ref{geneq} such that a unique local electromagnetic gauge transformation is mapped into tetrad vectors lying on the local future light cone. Since the ``point at infinity'' is the point on the asymptote of the upper branch of the hyperbola $X^{2}-Y^{2}=1$, then this point exists in the image of the map that we are studying. This is the point that maps into the north pole of the stereographic unit sphere. It is the electromagnetic gauge transformation that solved the inhomogeneous equation $D=1+C$ in sections \ref{diffeq}-\ref{diffeqrn}-\ref{geneq}. Analogous for the lower branch $X\geq0$ and $Y\leq0$ and also analogous for the opposite branch with $X\leq0$ and $Y\leq0$ and also $X\leq0$ and $Y\geq0$ with the unique solution for $D=-(1+C)$. The inhomogeneous solution for $D=-(1+C)$ is in the past light cone and corresponds to the point at infinity for the opposite branch of the hyperbola. Once we consider the point at infinity plus the upper and lower branches for both opposite branches of the hyperbola we close the curve at the north pole and we would have two closed curves. It is simple to see that we can also repeat the whole argument for the conjugate hyperbola $Y^{2}-X^{2}=1$ and its two branches. The conjugate hyperbola is a reflection of the original hyperbola through the asymptote $Y=X$. The reflected inhomogeneous solutions for the past and future light cones are the same as for the original hyperbola since they are invariant by reflection through the asymptote $Y=X$. By topological closure we would be able to map the conjugate hyperbola plus two infinities corresponding to $D=1+C$ and $D=-(1+C)$ and we will have two closed curves mapped into $SO(2)$. A total of four closed curves mapped into $SO(2)$ when also considering the conjugate hyperbola. The group LB1 is given by $SO(1,1) \times Z_{2} \times Z_{2}$ where $SO(1,1)$ is proper orthochronous. The first $Z_{2}$ is given by $\{I_{2 \times 2}, -I_{2 \times 2}\}$ and the second $Z_{2}$ is given by $\{I_{2 \times 2}, \mbox{the swap}\: (01|10)\}$. We would have to add in order to complete the image of the map $SO(1,1) \times Z_{2} \times Z_{2}\: \bigoplus \: \{light\:cone\:gauge\}$ where the light cone gauge includes the inhomogeneous two solutions to the differential equations in the local future and past light cones established in sections \ref{diffeq}-\ref{diffeqrn}-\ref{geneq} where the reflection through the asymptote $Y=X$ will produce two more identical inhomogeneous solutions. A total of four. We are suppressing the homogeneous solutions to these differential equations for the possible map of local gauge transformations into local tetrad transformations that take timelike and spacelike tetrad vectors on the local plane one into the intersection of the plane with the local light cone. These isomorphisms will be established modulo homogeneous solutions to the differential equations as found in sections \ref{diffeq}-\ref{diffeqrn}-\ref{geneq}. For more details see manuscript \cite{SING}.

\section{Appendix V}
\label{sec:appV}

We will study in this section how to make a suitable choice for the gauge vector $Y^{\alpha}$ for the Maxwell equations with a source $J^{\mu}$ as it is the case in a solenoid cylindrical geometry. Let us focus for practical purposes in the cylindrical problem as an example that permits a better visualization of this physical situation when the Maxwell equations have sources. The point is that in geometrodynamics, the Maxwell equations,

\begin{eqnarray}
f^{\mu\nu}_{\:\:\:\:\:;\nu} &=& J^{\mu} \label{L1M}\\
\ast f^{\mu\nu}_{\:\:\:\:\:;\nu} &=& 0 \ , \label{L2M}
\end{eqnarray}

tell us about the existence of one potential $X^{\alpha}=A^{\alpha}$. Then the question arises about gauging the vectors in equations (\ref{V3}-\ref{V4}). The tetrad of eigenvectors to the Einstein-Maxwell or Minkowski-Maxwell stress-energy tensor is given by,

\begin{eqnarray}
V_{(1)}^{\alpha} &=& \xi^{\alpha\lambda}\:\xi_{\rho\lambda}\:X^{\rho}
\label{V1NONFIXEDAPP}\\
V_{(2)}^{\alpha} &=& \sqrt{-Q/2} \:\: \xi^{\alpha\lambda} \: X_{\lambda}
\label{V2NONFIXEDAPP}\\
V_{(3)}^{\alpha} &=& \sqrt{-Q/2} \:\: \ast \xi^{\alpha\lambda} \: Y_{\lambda}
\label{V3NONFIXEDAPP}\\
V_{(4)}^{\alpha} &=& \ast \xi^{\alpha\lambda}\: \ast \xi_{\rho\lambda}
\:Y^{\rho}\ .\label{V4NONFIXEDAPP}
\end{eqnarray}

The non-zero components of the electromagnetic field are $f_{\theta\:z} = B_{\rho}$ and $f_{\rho\theta} = B_{z}$, with $A_{\theta}$ also given in reference \cite{NASA}. Then $\xi_{\rho\theta}=f_{\rho\theta}$ and $\xi_{\theta\:z}= f_{\theta\:z}$ in a flat Minkowskian spacetime with signature $(-+++)$ because the complexion is equal to zero. We also know that the metric in cylindrical coordinates $(t, \rho, \theta, z)$ will be diagonal $(-1, 1, \rho^{2}, 1)$. The metric determinant $g$ will satisfy $\sqrt{-g}=\rho$. For the alternating tensor cylindrical components, section IX, APPENDIX I in reference \cite{A} is useful when considering all these elements. It is also useful section \ref{sec:appI} in this manuscript. We notice that using the four-dimensional Lorentz flat Minkowski metric tensor in cylindrical coordinates allows the introduction at every point of four orthonormal vectors. Let them be $k^{\mu}_{t}=(1,0,0,0)$, $k^{\mu}_{\rho}=(0,1,0,0)$, $k^{\mu}_{\theta}=(0,0,\frac{1}{\rho},0)$ and $k^{\mu}_{z}=(0,0,0,1)$. Since the vector $k^{\mu}_{t}$ has non-trivial $t$ components, then we can choose the gauge vector $Y^{\alpha}=k^{\alpha}_{t}$. This way, the equation components (\ref{V3rhoEX}-\ref{V4tEX}) will not be trivial. In the Cylindrical geometry the only non-zero tetrad vector components for the local plane two will be,

\begin{eqnarray}
V_{(3)}^{\rho} &=& \ast \xi^{\rho\:t}\:Y_{t}=-\ast \xi_{t\rho}\: Y^{t} \label{V3rhoEX}\\
V_{(3)}^{z} &=&  \sqrt{Q/2} \:\:\ast \xi^{zt}\:Y_{t}=-\sqrt{Q/2} \:\:\ast \xi_{tz}\: Y^{t} \label{V3zEX} \\
V_{(4)}^{t} &=& \sqrt{Q/2} \:\:\ast \xi^{t\rho}\:\ast \xi_{t\rho}\:Y^{t} + \ast \xi^{tz}\:\ast \xi_{tz}\:Y^{t} = -\sqrt{Q/2} \:\:(\mid \ast \xi_{t\rho} \mid^{2} + \mid \ast \xi_{tz} \mid^{2})\:Y^{t} \ . \label{V4tEX}
\end{eqnarray}

where $Q = 2\:f_{\rho\theta}\:f^{\rho\theta}+2\:f_{\theta\:z}\:f^{\theta\:z}={2(B_{z}^{2} + B_{\rho}^{2}) \over \rho^{2}}$. In the Cylindrical geometry the only non-zero tetrad vector components for the local plane one will be,

\begin{eqnarray}
V_{(1)}^{\theta} &=& \xi^{\theta\rho}\:\xi_{\theta\rho}\:A^{\theta} + \xi^{\theta\:z}\:\xi_{\theta\:z}\:A^{\theta} \label{V1thetaEX}\\
V_{(2)}^{\rho} &=& \sqrt{Q/2} \:\:\xi^{\rho\theta}\:A_{\theta} \label{V2rhoEX}\\
V_{(2)}^{z} &=& \sqrt{Q/2} \:\:\xi^{z\theta}\:A_{\theta} \ .\label{V2zEX}
\end{eqnarray}

Despite the fact that in this cylindrical geometry we cannot choose the gauge vector $Y^{\alpha}$ to be $A^{\alpha}$ simply because the components of $V_{(3)}^{\alpha}$ and $V_{(4)}^{\alpha}$ will all be zero and nor we can choose $Y^{\alpha}$ to be $\ast A^{\alpha}$ simply because $\ast A^{\alpha}$ does not exist in the cylindrical geometry with source, we can choose it to be the vector $Y^{\alpha}=k^{\alpha}_{t}$. It is also possible to choose the following $Y^{\alpha}=k^{\alpha}_{t} + A^{\alpha}$ gauge-vector in plane two. In the term $k^{\alpha}_{t}$ there should be a constant that makes the units the same as in the term $A^{\alpha}$. We can then always choose another gauge by implementing $Y^{\alpha}=k^{\alpha}_{t} +A^{\alpha} \rightarrow Y^{\alpha}=k^{\alpha}_{t} + A^{\alpha} + \Lambda_{,\beta}\:g^{\alpha\beta}$. It is an electromagnetic potential gauge transformation for a valid choice of gauge-vector and we can study the tetrad eigenvector transformations in the local plane two exactly as in reference \cite{A} and section \ref{intro}. There would be no mathematical change in the analysis structure. The whole point of this section is to highlight that when we have in flat-Minkowski spacetime the Maxwell equations with sources, then we have to be careful with our choice for the gauge vector $Y^{\alpha}$. With this observation all the analysis about tetrad vector transformations in the local plane two will follow the same lines as in manuscript \cite{A} and the tetrad study originally made in section \ref{intro} for Einstein-Maxwell curved spacetimes without sources will stand. Let us remember that for Einstein-Maxwell curved spacetimes these steps were not necessary since there was a natural non-trivial choice $Y^{\alpha}=\ast A^{\alpha}$. In Einstein-Maxwell curved spacetimes with sources we would also have an analogous choice to the one given in this section depending on the case.

\section{Declaration of interest statement}
\label{interest}

The authors declare that they have no known competing financial interests or personal relationships that could have appeared to influence the work reported in this paper.

\section{Data availability statement}
\label{data}

There is no data to be reported in this paper.

%\bibliography{your-bib-file} % place the references here.

\end{document}